\documentclass[runningheads,a4paper,anonymous]{llncs}
\usepackage{llncsdoc}
\usepackage{color}
\usepackage{appendix}
\usepackage{float}
\usepackage{caption}
\usepackage{wrapfig} 
\usepackage{amssymb}
 \usepackage{graphicx}
\usepackage{multirow}
\usepackage{makecell}
\usepackage[margin=1.5in]{geometry}    
\usepackage[english]{babel}
\usepackage[utf8]{inputenc}
\usepackage{algorithm}
\usepackage{algpseudocode}
 \usepackage{amsmath}

\algnewcommand\algorithmicswitch{\textbf{switch}}
\algnewcommand\algorithmiccase{\textbf{case}}
\algnewcommand\algorithmicassert{\texttt{assert}}
\algnewcommand\Assert[1]{\State \algorithmicassert(#1)}%
\algdef{SE}[SWITCH]{Switch}{EndSwitch}[1]{\algorithmicswitch\ #1\ \algorithmicdo}{\algorithmicend\ \algorithmicswitch}%
\algdef{SE}[CASE]{Case}{EndCase}[1]{\algorithmiccase\ #1}{\algorithmicend\ \algorithmiccase}%
\algtext*{EndSwitch}%
\algtext*{EndCase}%

    \newenvironment{proofsketch}{%
  \proof}{\endproof}


\begin{document}
\title{Efficient Reactive Synthesis}
\author{ Xin Ye$^1$ and Harald Ruess$^2$
 }
\authorrunning{ }

\institute{$^1$fortiss - An Institut Technische Universit\"{a}t M\"{u}nchen\\
$^2$Entalus, FL, USA
}
\maketitle
\begin{abstract} 
Our main result is a polynomial time algorithm for deciding realizability for the GXU sublogic of linear temporal logic.
This logic is particularly suitable for the specification of 
embedded control systems, 
and it is more expressive than GR(1)\@. 
Reactive control programs for GXU specifications are 
represented as Mealy machines, which are extended by the monitoring  of input events. 
Now, realizability for GXU specifications is shown to be equivalent to solving a certain subclass of 2QBF satisfiability problems\@.
These logical problems can be solved in cubic time in the size of GXU specifications.
For unrealizable GXU specifications, 
stronger environment assumptions are mined from failed consistency checks
based on Padoa's characterization of definability
and Craig interpolation. 
\end{abstract}

\section{Introduction}

Structured requirements specification languages for embedded control systems such as EARS~\cite{mavin2009easy} and CLEAR~\cite{hall2018clear,hall2021knowledge} 
are semantically represented in a rule-like fragment of linear temporal logic (LTL)~\cite{rolls10227just}\@. 
Consider, for example, the EARS specification pattern
   \begin{align}\label{ex:ears}
       \text{When } T \text{ the system shall } A \text{ until } R\mbox{\@,}
   \end{align}
where constraint $T$ specifies the {\em trigger} condition, $A$ the {\em actions} to be triggered, and $R$ the corresponding {\em release} condition for these actions.  
Hereby, the constraints $T$, $A$, $R$ may contain any finite number of $\mathbf{X}$ ({\em next-step}), but no other, temporal operators. 
The logical meaning of~(\ref{ex:ears}) is easily encoded  by
    \begin{align}\label{ex:ears.ltl}
    \mathbf{G}\,(T \rightarrow (A \,\mathbf{U}\, R))\mbox{\@,}
    \end{align}
where $\mathbf{G}$ ({\em globally}) 
and $\mathbf{U}$ ({\em strong until}) are LTL temporal operators.
In this way, every EARS specification pattern
can be implemented in a rule-like sublogic 
of LTL without nested until operators~\cite{rolls10227just}\@.
We define a corresponding sublogic of LTL which we call GXU\@.

The GXU logic is of course motivated by the earlier GXW~\cite{CYR16,CLR17}, which has proven to be practically meaningful for modeling a large number of embedded control scenarios from industrial automation and for autogenerating correct-by-construction {\em programmable logic control} (PLC) programs~\cite{CYR16,xie2022structural}\@.
As the naming suggests, GXW only supports {\em weak until}, whereas GXU also includes the {\em strong until} operator\@.
With this extension GXU can now express all
GR(1) specifications~\cite{bloem2012synthesis}\@. 

We consider {\em Church's synthesis problem}, which involves generating an input-output control program to implement a given control specification.  
Solutions to Church's problem
are traditionally based on the fundamental {\em logic-automaton
connection}, and they rely on rather complex determination procedures and emptiness tests for $\omega$-automata \cite{buchi1990solving,rabin1969decidability,abadi1989realizable,pnueli1989synthesis,bohy2012acacia+,filiot2011antichains,jobstmann2006optimizations,faymonville2017encodings,finkbeiner2012lazy,jobstmann2007anzu,schewe2007bounded,kupferman2005safraless}\@.
However, in the special case of GXU with its syntactic restrictions, we can easily compile specifications to programs as long as the individual specification rules are mutually consistent.
More precisely, this {\em structural} 
approach~\cite{cheng2016structural} to GXU reactive synthesis
proceeds in two subsequent steps by:
\begin{enumerate}
    \item \label{step1} Recursing on the syntactic structure of the GXU control 
          specification for instantiating and combining a small set of GXU-specific {\em Mealy machines with monitors} templates. 
          The transitions of the resulting Mealy machines are equipped with automata that monitor finite windows of input traces to determine when trigger conditions exist.
   \item \label{step2}
   Determining possible conflicts between individual GXU constraints by solving a corresponding validity problem. 
\end{enumerate}
Consistency checks in (step~\ref{step2}.) are naturally encoded as {\em forall-exists} top-level quantified Boolean formulas (2QBF)~\cite{CYR16}\@.  
Due to the syntactic restrictions of GXU we are able to
solve this specific class of logical problems in polynomial time.
The complexity of consistency checking (step \ref{step2}.) dominates over the structural recursion (step~\ref{step1}\@.)\@,
and we obtain our main result that realizability for GXU is decidable in polynomial time.

In comparison, the synthesis procedure for GXW in~\cite{CYR16} is incomplete as it generally does not deal with interdependent specifications, and GR(1) synthesis
is polynomial in the underlying state space, which itself can be exponential.
Furthermore, since GXU synthesis is essentially based on logical operations, it can easily be extended to support a richer control specification language that includes 
clocks and other decidable data constraints. 

Specifying the operating environment of a controller arguably is one of the most challenging steps in designing embedded control systems. 
For example, the GXU specification
     \begin{align}\label{ex:assumption}
     \mathbf{G}(\mathrm{a} \rightarrow \mathbf{X}\, \mathrm{b}) ~\wedge~
         \mathbf{G}(\mathrm{b} \rightarrow \mathbf{X} \neg \mathrm{b})\mbox{\@,} 
      \end{align}
 clearly is not realizable. Whenever the input event '$\mathrm{a}$' holds in two consecutive steps, the 
system can not produce a value for the output event '$\mathrm{b}$' such 
that specification~(\ref{ex:assumption}) holds.
A sensible repair to~(\ref{ex:assumption}) therefore is to constrain input events by the temporal constraint
$\mathbf{G}(\mathrm{a} \rightarrow \mathbf{X} \neg \mathrm{a})$\@.
    
The general problem of {\em assumption mining} is to find environment restrictions which are sufficient for making the system under consideration realizable, while still giving the environment maximal freedom~\cite{CHJ08}\@.
Assumption mining solutions constrain the behavior of the environment by excluding countertraces that come from failed synthesis attempts~\cite{CHJ08,cheng2014g4ltl,alur2015pattern,cavezza2017interpolation,cavezza2020minimal,alrajeh2020adapting,dimitrova2019synthesizing,baumeister2020explainable,li2011mining}\@.
Since we have reduced the realizability problem for GXU to a logic problem, we are now able to mine environmental assumptions in a symbolic manner, which solely relies on a combination of logical notions, including
{\em Padoa's} characterization of definability, 
and the corresponding construction of {\em interpolation} formulas for restricting environmental behavior.

Overall, our main contributions are as follows.
\begin{enumerate}
     \item Definition of the GXU logic, which is more expressive than both GXW~\cite{CYR16,CLR17} 
     and GR(1)~\cite{bloem2012synthesis}\@. 
     \item Definition of the concept and use of {\em Mealy machines with monitors}, where each transition is equipped with an interleaved automaton that monitors the input events and is used to determine the generation of the corresponding output events\@. 
     \item A polynomial time synthesis method for GXU specifications based on a reduction of the realizability problem to an equivalent and polynomial time validity problem.  
     \item Assumption mining based on a combination of logic-based concepts such as Padoa's theorem 
     and Craig interpolation.
\end{enumerate}
The paper is structured as follows.
Section~\ref{sec:prelims} sets the stage by summarizing fundamental logical concepts and notation. 
Next, Section~\ref{sec:gxu} defines the GXU sublogic of LTL, and
Section~\ref{sec:mealymachine}
describes the structural approach to construct, 
from a small finite amount of patterns, 
a Mealy machine with monitors for realizing a single GXU formula.
Synthesis for GXU is described in Section~\ref{sec:synthesis}, including (1) the use of Mealy machines to represent streams satisfying the given GXU formula, and (2) a reduction of GXU realizability to a logical validity problem.
In Section~\ref{sec:repair} we describe a logical approach to assumption mining for GXU.
We illustrate the reactive synthesis and repair of GXU specifications in Section~\ref{sec:case.study} using a case study from the field of automation. 
In Section~\ref{sec:related} we compare these results with the most closely related work, 
and Section~\ref{sec:conclusion} concludes with a few remarks. 
\label{sec:intro}
\section{Preliminaries}

We review some basic concepts and notation of propositional and temporal logic as used throughout this exposition.
  
\paragraph{Propositional Logic.}
A {\em propositional formula} is built from
propositional variables in the given set $V$ of variables and the usual propositional connectives $\neg$, $\wedge$, $\vee$, $\leftrightarrow$, and $\rightarrow$\@.
The set of variables occurring in a formula $\varphi$ is denoted by $var(\varphi)$\@.
\emph{Literals}, \emph{terms}, \emph{clauses}, 
\emph{conjunctive (disjunctive) normal forms}, 
\emph{variable assignments},
and \emph{(un)satisfiability} of formulas are defined in the usual way. 
{\em Craig interpolation} states that for all propositional formulas $\varphi$, $\psi$ such that 
  $\varphi \wedge \psi$ is unsatisfiable,
  there is an {\em interpolant} $I$ for $\varphi$, $\psi$\@; that is: (1) $\varphi$ implies $I$,
  (2) $I \wedge \psi$ is unsatisfiable, and
  (3) $vars(I) \subseteq vars(\varphi) \cap vars(\psi)$\@.

  Let $\varphi$ be a propositional formula and $X \subseteq vars(\varphi)$\@. 
  A variable $x$ is 
  {\em defined} in terms of $X$ in $\varphi$ if $\sigma(x) = \tau(x)$ for any two satisfying variable assignments $\sigma$ and  $\tau$ of $\varphi$ that agree on $X$\@. 
  A {\em definition} of a variable $x$ by $X$ in $\varphi$ is a formula $\psi$ with $vars(\psi) \subseteq X$  such that  $\sigma(x) = \sigma(\psi)$ for any satisfying assignments $\sigma$ of $\varphi$\@.
  Padoa's characterization of definability can be used to determine whether or not a certain variable is defined~\cite{padoa1901essai,lang2008propositional}\@.
  \begin{lemma}[Padoa~\cite{padoa1901essai}]\label{padoa}
  Let $\varphi$ be a propositional formula, 
  $X \subsetneq vars(\varphi)$, 
  $v \in vars(\varphi) \setminus X$\@, and
  $\varphi'$ be the propositional formula obtained by replacing every variable $y \in vars(\varphi) \setminus X$ with a fresh variable $y'$\@.
  Then $v$ is defined by $\varphi$ in $X$ if and only if the formula $\varphi \wedge v \wedge \varphi' \wedge \neg v'$ is unsatisfiable.
  \end{lemma}
  Now, a definition for variable $v$ is obtained as a {\em Craig interpolant} of an unsatisfiable conjunct $(\varphi \wedge v) \wedge (\varphi' \wedge \neg v')$. 
  Such a definition can be efficiently extracted from a proof of definability 
  based on, say, a  SAT solver that generates resolution proofs~\cite{krajivcek1997interpolation}\@.

\paragraph{Quantified Propositional Logic.}
A {\em quantified Boolean formula} (QBF) includes, in addition, {\em universal} ($\forall$) and {\em existential} ($\exists$) quantification over variables in $V$\@. 
Every QBF formula can be converted into {\em Skolem normal formal} 
      $
  (\exists f_1, \ldots, f_n)(\forall y_1, \ldots, y_m)\,\varphi$\@, 
  where $\varphi$ is a propositional formula, 
  by systematically "moving" an existential quantification before a universal, based on the equivalence of the QBF $(\forall x)(\exists y)\,\varphi(x, y)$ with
  $(\exists f)(\forall x)\varphi(x, f(x))$\@. 
  Hereby, the existential quantification on $f$ is interpreted over functions $\{0, 1\}^k \to \{0, 1\}$, where $k$ denotes the number of dependencies on universally quantified variables\@.
  The functions $f$ are referred to as {\em Skolem functions}\@. 

\paragraph{Linear Temporal Logic.}
For a set $V$ of variable names,  formulas in LTL are constructed inductively from 
the constants $\mathit{true}$, $\mathit{false}$, 
variables in $V$, 
the unary negation operator $\neg$, the binary \emph{strong until} operator $\textbf{U}$, 
and the unary \emph{next-step} operator $\textbf{X}$\@.
The semantics of LTL formulas is defined with respect to $\omega$-infinite traces of variables $V$ in the usual way. 
An LTL formula of the form $\varphi_1\, \textbf{U}\, \varphi_2$, for example, holds with respect to an $\omega$-infinite trace if $\varphi_1$ holds along a finite prefix of this trace until $\varphi_2$ holds. 
$w\models \varphi$ is used to denote that $\varphi$ holds for the trace $w$\@. 
We also make use of defined temporal operators, including $\varphi \wedge \psi = \neg(\varphi \vee \neg \psi)$, $\varphi \rightarrow \psi = \neg \varphi \vee \psi$, $\textbf{F} \varphi = \mathit{true} \,\textbf{U}\, \varphi$ and $\textbf{G} \varphi = \neg (\mathit{true} \,\textbf{U}\, \neg \varphi)$\@. 
$L(\varphi) = \{ w\in  2^{V}| w\models \varphi\}$ denotes the language associated with an LTL formula $\varphi$\@. 
 
%
%
%

Now, assume the variables $V$ to be partitioned into $V_{in}$ and $V_{out}$, which are interpreted over some finite domain (of events)\@. 
Intuitively $V_{in}$ is the set of {\em input variables} controlled
by the environment, and $V_{out}$ is the set of {\em output variables} controlled by the system. 
An {\em interaction} $(x_0,y_0),(x_1,y_1),\cdots$ in $( V_{in}\times V_{out})^{\omega}$   an infinite sequence of pairs of input and corresponding output events. 
Such an interaction is produced by a {\em synchronous program} $P_s:V_{in}^+ \to V_{out}$, which determines output $y_i$ given the input history $x_0\ldots x_i$ for all time steps $i$\@.
For example, every Mealy machine determines a synchronous program. 
If the LTL formula $\varphi(V_{in};V_{out})$ holds for the interactions produced by $P_s$, then $P_s$ is said to {\em (synchronously) realize} $\varphi(V_{in}; V_{out})$\@.
Moreover, $\varphi(V_{in};V_{out})$ is said to be {\em (synchronously) realizable} if there exists such a synchronous program $P_s$~\cite{klein2012effective}\@. 

For specifications written in propositional LTL, the worst case complexity of the realizability problem is doubly exponential~\cite{pnueli1990distributed}\@.
More efficient algorithms exist for fragments of LTL such 
as {\em Generalized Reactivity(1)}~(GR(1))~\cite{piterman2006synthesis}, which consists of LTL formulas of the form
    $
        \textbf{GF}\;\varphi^0_1 \wedge \ldots \wedge \textbf{GF}\;\varphi^0_m
      \rightarrow \textbf{GF}\;\psi^0_1 \wedge \ldots \wedge \textbf{GF}\;\psi^0_n\mbox{\@,} 
$
where each $\varphi^0_i$, $\psi^0_j$ is a Boolean combination of atomic propositions. 
Realizability for GR(1) is solved in time $N^3$, 
where $N$ is the size of some underlying 
state space~\cite{piterman2006synthesis}\@.


\label{sec:prelims}

\newcommand{\GXU}{{GXU}}
\newcommand{\GXW}{{GXW}}
\newcommand{\GRone}{{GR}(1)}

\section{GXU Logic}

%
In the definition of GXU logic
we take a set $V$ of variables as given,
which is partitioned into the set of
{\em input variables} $V_{in}$
and {\em output variables} $V_{out}$\@.
The letter $p$ is used to denote a literal of the form $v$ or $\neg v$, for $v\in V$\@, 
and $\phi^0$ denotes a propositional formula without any temporal operators.
Furthermore, an LTL formula is said to be {\em basic up to $i$}, for $i \geq 0$,
if it is a propositional combination with literals of the form
          $\textbf{X}^j p$, 
          $\neg \textbf{X}^j p$, for $0 \leq j\leq i$\@.\footnote{$X^0 \psi$ is defined to be $\psi$\@.}
\begin{definition}[GXU Formulas]\label{def:gxu.formulas}~\\
A {\em GXU formula} (in the variables $V$) is an LTL formula of the form (1)-(4)\@. 
\begin{enumerate}
\item ({\em Reaction}) $\textbf{G}(\varphi \rightarrow \textbf{X}^i \psi)$, $i \geq 0$,
with the following restrictions: $\varphi$ is basic up to $i$,
    \begin{enumerate}
    \item  $\psi$ is restricted to be a literal $p$\@. \label{lbl:one-a}
    \item $\psi$ is of the form $(p\;\textbf{U}\;\psi')$ with $p$ a literal and  $\psi'$ a basic formula up to $i$\@. \label{lbl:one-b}
    \end{enumerate}

\item\label{lbl:three} ({\em Invariance})  
     $\textbf{G}(\varphi \leftrightarrow \textbf{X}^i p)$,
     where $p$ is a literal and $\varphi$ 
     is basic up to $i$\@. 

\item ({\em Global Invariance}) $\textbf{G}\,\phi^0$\@. 

\item ({\em Liveness})  $\textbf{F}\,\phi^0$\@.\label{lbl:four}
\end{enumerate}
\end{definition}
GXU logic therefore extends GXW~\cite{CLR17,CYR16}
by additionally supporting the {\em strong until} ($\mathbf{U}$) and the {\em eventually} ($\mathbf{F}$) temporal operators.

 
\begin{definition}[Reactive GXU Specifications]\label{GXU.spec}~\\
A {\em (reactive) GXU specification} (in the variables $V = V_{in} \cup V_{out}$) is of the form
   \begin{align}
   E \rightarrow S\mbox{\@,}
   \end{align}
where the {\em environment assumptions} $E$ are a finite (and possibly empty) conjunction of GXU formulas in the inputs $V_{in}$, and the {\em control guarantees} $S$ are a finite conjunction of GXU formulas in both the inputs $V_{in}$ and the outputs $V_{out}$\@. 
\end{definition}
A GXU reactive specification is also said to be in {\em assume-guarantee} format. 
\noindent
We illustrate the expressive power of GXU logic with some simple examples.

\begin{example}[Automatic Sliding Door~\cite{CHJ08}]
It is assumed that the input $in_0$ holds when someone enters the detection field, 
$in_2$ holds when the door reaches the end,
and $out$ holds when the opening motor is on. 
Now, a GXU specification of an automatic sliding door contains rule-like GXU formulas such as
$
\textbf{G}(\neg in_0 \wedge \textbf{X} in_0 \rightarrow \textbf{X}(out \;\textbf{U}\; in_2))\mbox{\@.}$
\label{ex:door}
\end{example}
The GXU formula~(in Example \ref{ex:door}) contains a combination of strong until and next-step operators,
which is not expressible in GR(1)\@. 
On the other hand, all GR(1) subformulas of the form $\textbf{GF}\;\varphi^0$ (see Section~\ref{sec:prelims})
can be expressed in terms of an equivalent \emph{GXU} specification\@.  
This is easy to see as a GR(1) subformula $\textbf{GF}\;\varphi^0$  is equivalent to the GXU reactive pattern
$\textbf{G}(\mathit{true} \rightarrow (\mathit{true} \;\textbf{U}\; \varphi^0))$\@.
\begin{lemma}
	GXU specifications
 are strictly more expressive than GR(1)\@.
\end{lemma}

\begin{example}[Hamiltonian paths]
A path in a graph is \emph{Hamiltonian} if it visits each vertex exactly once.  
For each vertex $v$ in the graph introduce two  variables: (1) $c_v$ for the currently visited vertex and (2) $c'_v$ for the vertex which is chosen to be the visited next. Now, the Hamiltonian path condition is expressed as the GXU formula $\varphi_1 \wedge \varphi_2 \wedge \varphi_3$ where $\varphi_1 $ is a conjunction of invariance formulas for picking the successors. 
That is,  for each edge $(v_1,v_2)$, we have $\textbf{G}(c'_{v_2} \leftrightarrow \textbf{X} c_{v_2})$, 
$\varphi_2$ is a conjunction of GXU liveness formulas to state that every vertex occurs at most once; that is, $\textbf{F}c_v$ for each vertex $v$, and $\varphi_3 $ is a GXU global invariance formula to show the edges;  that is,  for each edge $(v_1,v_2)$, we have  $c_{v_1}\wedge c'_{v_2}$.
Clearly, this encoding of the Hamiltonian path condition in terms of a GXU formula is exponential. 
\end{example}

\paragraph{Problem Statement.}
Given a \emph{GXU} specification  $E \rightarrow S$ in the input-output variables
$V = V_{in} \cup V_{out}$, the problem of
{\em GXU reactive synthesis}
is the construction of a synchronous program $P_s:  V_{in}^+ \to V_{out}$, represented as a Mealy machine, 
so that $P_s$ realizes $E \rightarrow S$\@.
The construction of a synchronous program to implement a given GXU
specification $\mathit{Spec} := E \rightarrow  S$
takes place in two consecutive steps: (1) compositional construction of Mealy machines extended with monitor automata from the \emph{GXU} control guarantees $S$; and (2) construction of a finite set of 2QBF validity problems for checking the consistency, and therefore realizability of $\mathit{Spec}$, of the $S$ control guarantees relative to the environment assumptions $E$\@.
    
If the reactive specification $\mathit{Spec}$ is unrealizable, stronger assumptions than $E$
about the control environment
are synthesized to rule out identified causes of unrealizability.


\label{sec:gxu}

\section{Mealy Machine with Monitors}
\label{sec:mealymachine}

The translation of a single GXU formula $\varphi$ into a realizing synchronous program $P$ is adapted from 
{\em structural synthesis}~\cite{CYR16}\@. 
The main difference is the translation of {\em strong until} in GXU and the use of Mealy machines extended with monitoring automata, instead of synchronous data flow programs as in~\cite{CLR17,CYR16}, to represent synchronous programs. 

 \subsection{Representing Reactive Programs.}
An \emph{$l$-length monitor} is a tuple $\mathcal{M}=(Q,F,\Sigma,\Delta,\theta)$ where $Q$ is the set of states, $F\subseteq Q$ is the set of final states, $\Sigma$ is the alphabet, $\Delta:Q\times \Sigma \rightarrow Q$ is the transition function, and $\theta:\Sigma^{N}\rightarrow \{\text{true},\text{false}\}$ is said to be a {\em word valuation}\@.  A \emph{run} $\pi=q_{0}q_{1}\cdots q_{l-1}$ of the monitor $\mathcal{M}$ on a word $w$ is a sequence of states such that for $i\in\{1,\cdots,l\},(q_{i-1},w[i-1],q_{i})\in \Delta$ and $w=w[0]w[1]\cdots w[l]$. A word $w$ of length $l$ is accepted if it  reaches a final state such that the word valuation $\theta(w)=\text{true}$\@. 
\begin{figure}[t] 
 \begin{center}
 	\includegraphics[width=8cm]{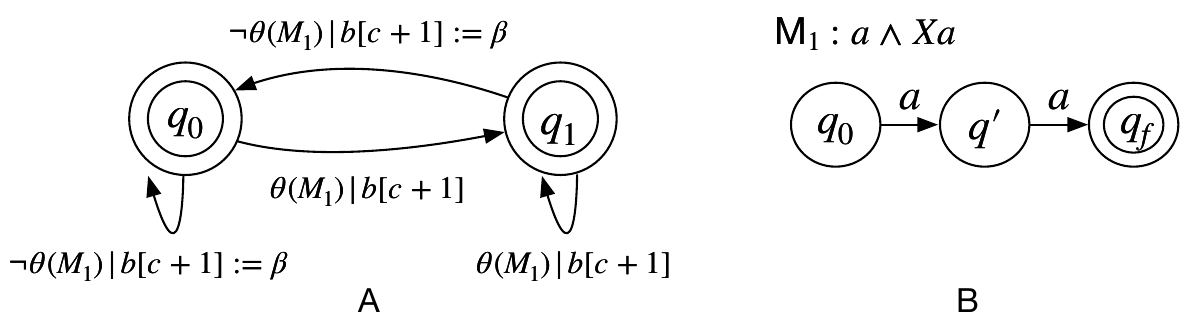}
  \caption{A: Mealy machine for $\textbf{G}(a\wedge \textbf{X}a \rightarrow \textbf{X} b)$, B: monitoring automata $\mathcal{M}_1$ for $a\wedge \textbf{X}a$  }
  \label{fig:fmealymachine}
 \end{center}
\end{figure}

  A \emph{placeholder} $\beta$ is an assignment  for each variable  used to denote all possible values.  Given a GXU formula $\varphi$ over a set of variables and a run $\pi$ with a set of assignments of variables to placeholders, replacing the placeholders with any of assignments results that $\varphi$ holds in the run $\pi$. Taking $\varphi=\textbf{G}(a\rightarrow b)$ for example, if $a$ doesn't hold, $\varphi$ holds no matter what the value of $b$ is. In this case, the placeholder $\beta$ is used to indicate the value of $b$ such that $(\neg a,\beta),(\neg a,\beta),(a,b)$ satisfies $\varphi$ where replacing $\beta$ with any assignment makes $\varphi$ holds\@.
  
\begin{definition}[Mealy Machines with Monitors]\label{def:M3}~\\ 
A \emph{Mealy machine $\mathcal{T}$ with monitors} is a tuple 
    $(\Sigma,Q, F',\Theta,\delta,q_0,\sigma)$\@,
    where $\Sigma$ is the alphabet, 
    $Q$ is the set of states, $F'\subseteq Q$ is the accepting state, 
    $q_0 \in Q$ is the initial state, 
   { $\Theta$ is the set of  word valuations}, 
    $\delta:Q\times \Theta\times (\Sigma\cup \{\beta\})^*    \rightarrow Q\times (\Sigma \cup  \{\beta\})$
is the state transition function\@,
which is activated if a word valuation holds on the input event based on the word valuation of \emph{monitors}, and $\sigma :Q\times  (\Sigma\cup \{\beta\})^* \rightarrow\Sigma \cup  \{\beta\} $ is the output function.
Instead of $\delta(q,\theta, w) = (q', o)$ we also write $q \xrightarrow{w,\theta,o} q'$\@.
\end{definition}
Note that Mealy machines with monitors are closely related to the concept of {\em nested weighted automata}~\cite{chatterjee2016quantitative},
and also to Mealy machines with B\"uchi acceptance~\cite{kini2015limit}.

The operational semantics of a Mealy machine $\mathcal{T}$ with monitors is based on configurations of the form
$(q,\langle w_1,w_2 \rangle)$,
for $q\in Q$ a state, 
$w_1 \in \Gamma^*$ an input, and 
$w_2 \in \Gamma^*$ a corresponding output word.
The machine ${\cal T}$ transitions by
   $q \xrightarrow{w,\theta_{\mathcal{M}},o} q'$
from state $q$ to $q'$ and produces the output $o$ only if the word valuation of the monitor $\mathcal{M}$ for $w$ holds\@.
For each $n$ such that
$0\leq n <|w|-1$, 
let $w'$ be a word with $w'[n]=\beta$ and $w'[|w|-1]=o$\@.
With this notation, the transition relation
	  $(q,\langle w_1,w_2 \rangle)\Rightarrow_{\mathcal{T}}(q',\langle w_1w,w_2 w' \rangle$
on configurations is defined to hold
if and only if $q  \xrightarrow{w,\theta_{\mathcal{M}},o} q'$\@.
As usual, $\Rightarrow_{\mathcal{T}}^i$, for $i \in \mathbb{N}$, and  $\Rightarrow_{\mathcal{T}}^*$ denote respectively the $i$-step 
iteration and the reflexive-transitive closure of the transition relation
$\Rightarrow_{\mathcal{T}}$ on configurations\@.

\begin{example}
Consider the Mealy machine $\mathcal{T}$ with monitors in Figure~\ref{fig:fmealymachine}(A)\@.
The definition of this machine $\mathcal{T}$ is based on
the $2$-length monitor $\mathcal{M}$ for the trigger condition $a\wedge \textbf{X}a$, as shown in 
Figure~\ref{fig:fmealymachine}(B)\@.
For the input $w=aa\neg aaa\ldots$, $\mathcal{T}$ checks whether the guard holds or not via the monitor $\mathcal{M}$ by the word evaluation of a $2$-length of sub-word at each time point. At time point $0$, the $2$-length sub-word $w[0,1] = aa$\@. 
Since the monitor $\mathcal{M}$ is accepting i.e. $\theta(w[0,1])=\text{true}$, $\mathcal{T}$ moves $q_0$ to $q_1$ and it produces the output $w'[0]=\beta,w'[1]=b$ with configuration $(q_1,\langle w[0,1],w'[0,1] \rangle)$\@. 
Now, $\mathcal{T}$ is at state $q_1$ and $\mathcal{M}$ monitors $w[1,2] =a\neg a$\@. 
$\mathcal{M}$ does not accept, and therefore $\mathcal{T}$ moves to state $q_0$ producing the output $w'[2]=\beta$\@. 
If we continue this process, we obtain $w'[3]=\beta$ and $w'[4]=b$\@,
and therefore the interaction $(a,\beta),(a,b),(\neg a,\beta),(a,\beta),(a,b),\cdots$\@. 
In this way, the machine $\mathcal{T}$
in~Figure~\ref{fig:fmealymachine}(A)
is shown to realize the GXU formula
   $G(a \wedge \textbf{X}a \rightarrow  \textbf{X} b)$\@.
\end{example}

 \subsection{Realization of GXU formulas} 
 \label{sec:realization}
 For a given GXU formula we construct a realizing Mealy machine with monitors by closely 
 following the case distinction in
 the Definition~\ref{def:gxu.formulas} of GXU formulas\@. 
 \begin{figure}[t] 
 \begin{center}
 	\includegraphics[width=13cm]{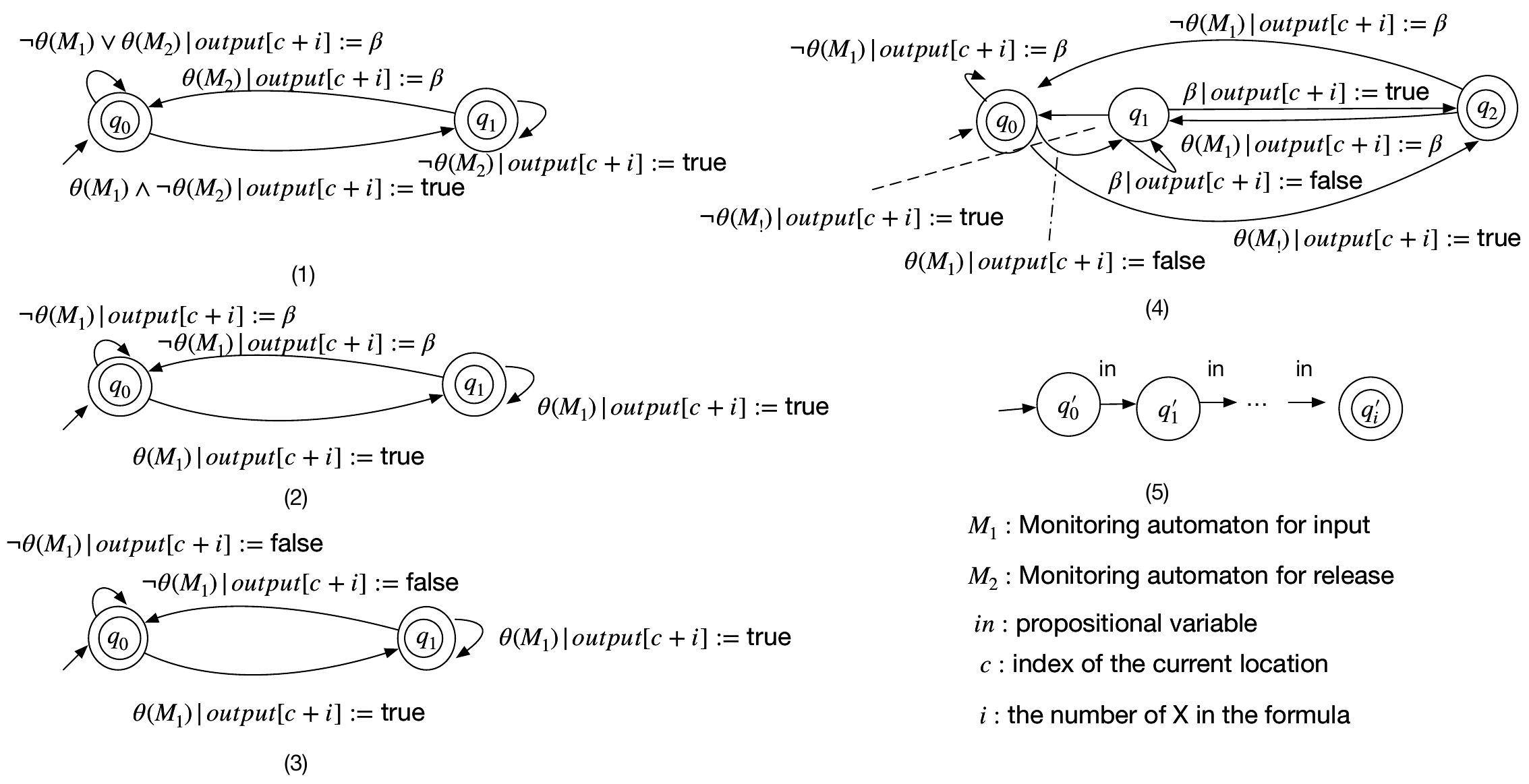}
  \caption{Templates of Mealy machine with monitors for the GXU formula patterns.}
  \label{fig:mealymachine}
 \end{center}
\end{figure}
 Let's start by looking at the \emph{reactions}~(\ref{lbl:one-a}) 
 in Definition~\ref{def:gxu.formulas}
 of the form $\textbf{G}(\varphi\rightarrow \textbf{X}^i \psi)$\@. 
 The construction of a monitor $\mathcal{M}$ for the trigger condition $\varphi$  corresponds to building a deterministic finite automaton from a regular expression.
 This trigger condition needs to be checked at every step, and therefore it relies on the word valuation of a monitor $\theta(w)$, 
 and the  Mealy machine in Figure~\ref{fig:mealymachine}(2) for the reaction pattern above introduces a special state for indicating whether the trigger condition $\varphi$ holds or not. Initially, this Mealy machine is at state $q_0$\@. If the valuation of the subtrace $w_1$ for the monitor $\mathcal{M}$ holds, that is $\theta_{\mathcal{M}}(w_1)$ holds, then it transitions from state $q_0$ to $q_1$ and it generates the  $\mathit{true}$ output\@. 
  Otherwise, 
  in case $\neg{\theta_{\mathcal{M}}(w_1)}$ holds, 
  it stays in state $q_0$ and produces the placeholder $\beta$ output. 
  Both $q_0$ and $q_1$ are accepting.  

GXU reactions~(\ref{lbl:one-b}) of the form $\textbf{G}(\varphi \rightarrow \textbf{X}^i (g \textbf{U} \psi'))$ are equivalent to the conjunction 
  \begin{align}\label{encoding.strong.until}
   \textbf{G}(\varphi \rightarrow \textbf{X}^i( g \textbf{W} \psi')) &\wedge \textbf{G}(\varphi \rightarrow \textbf{X}^i  ( \textbf{F} \psi'))\mbox{\@.}   
  \end{align}
The corresponding construction in Figure~\ref{fig:mealymachine} therefore 
includes two Mealy machines $\mathcal{T}_1$ (Figure~\ref{fig:mealymachine}~(1)) and $\mathcal{T}_2$ (Figure~\ref{fig:mealymachine}~(4)), so that  $\mathcal{T}_1$ implements the left conjunct of~formula (\ref{encoding.strong.until}) 
$\mathcal{T}_1$ implements the corresponding right conjunct, 
 and $\mathcal{T}_1$ and $\mathcal{T}_2$ share
 both the monitor and the word valuation. 
 The additional state $q_2$ indicates that the
 trigger condition holds and also that $\psi'$ holds.
 Finally, there are transition from $q_1$ (and $q_2$) back to $q_0$ whenever the word valuation of the current subtrace $w_1$ does not hold;  that is, $\neg{\theta_{\mathcal{M}}(w_1)}$ holds\@. In $\mathcal{T}_2$, $q_0$ and $q_2$ are accepting states.

%

The construction for \emph{invariance} formulas
is similar to  the one for \emph{reactions},
but additional care is required when handling this equivalence and the output variables.
And, finally, the case of \emph{liveness} formulas is reduced to the one for the reactions $(\mathit{true}\rightarrow \textbf{F}p)$\@. 

Overall, the Mealy machines with monitors that realize the GXU formulas in Definition~\ref{def:gxu.formulas}
are summarized in 
Figure~\ref{fig:mealymachine}\@.
Thereby, the machines in Figure~\ref{fig:mealymachine}(1) 
and~\ref{fig:mealymachine}(4) realize
reactions $\textbf{G}(\varphi \rightarrow \textbf{X}^i (g \textbf{U} \psi'))$\@;
the machines  in Figure \ref{fig:mealymachine}(2)
realize reactions $\textbf{G}(\varphi \rightarrow \textbf{X}^i \psi')$\@;
Figure~\ref{fig:mealymachine}(3) displays the machine  for realizing invariance formulas; and Figure~\ref{fig:mealymachine}(4) is adapted to represent the machine for
realizing liveness formulas.

The machine constructions in this section thus lead to a reactive program $P$, represented as a Mealy machine with monitors, for realizing a given GXU formula $\varphi$\@.
The basic machine templates in Figure~\ref{fig:mealymachine} 
are used to construct control programs
from their specifications.




\section{GXU Synthesis}

Let's consider as given a GXU specification $\mathit{Spec} := E \rightarrow S$ with variables in
$V = V_{in} \cup V_{out}$ (Definition~\ref{GXU.spec})\@. 
Controllers for realizing $\mathit{Spec}$
are represented as a set $\mathcal{T}_S$ 
of Mealy machines with monitor, as obtained from the syntactic translations in Section~\ref{sec:mealymachine} 
of the control guarantees $S$\@.
We now construct a 2QBF formula to characterize that for all possible inputs that satisfy the given environment assumptions, there is at least one output that satisfies the associated control guarantees.
For this purpose, the environment assumptions $E$ are also translated into a set 
$\mathcal{T}_E$
of Mealy machines with monitors. 
In this way, realizability for $\mathit{Spec}$ is reduced to the validity problem for a certain subclass of 2QBF formulas\@.
The main development of this section is a polynomial time {\em consistency check} for this subclass of formulas\@.

\subsection{Consistency Check}


\paragraph{Completeness Bound.} 
{Completeness thresholds} are used 
 in bounded model checking to determine a bound $k$
 so that if no counterexample of length $k$ or less is found for a given LTL formula then the formula holds for all 
 paths in this model.   

\begin{definition}[Linear Completeness Threshold]
\begin{itemize}
\item  A {\em linear completeness threshold $k$} for a GXU formula $\phi$ is a linear (in the size of 
$\phi$) bound such that there exists a Mealy machine $\mathcal{T}$, $0\leq j \leq k, c_0 \Rightarrow_{\mathcal{T}}^*(q,\langle w[0,j],w'[0,j]\rangle)$ with $ q\in F$, $c_0$ is the initial configuration and $(w[0],w'[0])(w[1],w'[1])\cdots(w[j],w'[j])\models \phi$, then $\omega,\omega'\in \Gamma^*,c_0\Rightarrow_{\mathcal{T}}^*(q_f,\langle \omega,\omega'\rangle), q_f \in F$ such that $(\omega[0],\omega'[0])(\omega[1],\omega'[1])\cdots\models \phi$. 
\item For a given GXU specification $E\rightarrow S \text{ where } E=\phi_1\wedge \phi_2\wedge \cdots \wedge \phi_n, S=\phi_1\wedge \phi_2\wedge \cdots \wedge \phi_n'$, a linear completeness threshold $K$ is a smallest linear bound such that $k_1,\cdots,k_n,k'_1,\cdots, k'_{n'}\leq K$ where $ k_1,\cdots,k_n,k'_1,\cdots, k'_{n'}$ are linear completeness thresholds for sub-formulas. 
 \end{itemize}
 \end{definition}
 
\begin{lemma}
	Every GXU formula and every GXU specification 
    has a linear completeness threshold. These thresholds can all be computed in linear time. 
\end{lemma}
\begin{proofsketch}
   The directed graphs of  the Mealy machines in Figure~\ref{fig:mealymachine} for
   realizing GXU formulas 
   are all {\em cliquey}, that is, 
   every maximal strongly connected component is a bidirectional clique\@. 
   In particular, every node has a self-loop.
   The claim follows from the fact 
    that every cliquey generalised B\"{u}chi automaton has a linear completeness threshold~\cite{lct}\@.   That is, every Mealy machine $\mathcal{T}$ as constructed from a GXU formula $\phi$ has a linear completeness threshold, say $k_{\phi}^{\mathcal{T}}$  constructed as shown in Table~1 in 
  the Appendix~\ref{appendix:completeness} in linear time\@. Let $l$ be the length of the associated monitor for $\phi$, then the completeness threshold $k_{\phi}$ for $\phi$ equals $l\times k_{\phi}^{\mathcal{T}}$. 
  Let $\bigwedge_{0\leq m'_{E}\leq m}\varphi_{m'_{E}} \rightarrow\bigwedge_{0\leq m'_{S}\leq m'}\psi_{m'_{S}}$ be the given GXU specification\@.
Hence, the completeness threshold for the given GXU specification $K$ is  $\mathit{max}(k_{\varphi_{0}},\cdots, k_{\varphi_{m}},  k_{\psi_{0}},\cdots, k_{\psi_{m'}})$.
\end{proofsketch}

\paragraph{Encoding.}  Based on the existence of a linear completeness
threshold $K$, we  formulate a complete consistency 
check for a given GXU specification $\mathit{Spec}$.
We introduce, for $0\leq j \leq K$, new sets of (input) variables 
   $V_{in}[j] := \{ x[j]\,|\, x\in V_{in}\}$
and (output) variables
   $V_{out}[j] := \{ y[j]\,|\, y\in V_{out}\}$\@.
These variables are used to specify the values of input and output variables at transition step $j$\@.
The process of {\em encoding} makes these pointwise Boolean values explicit.
For example, the encoding of the formula  $\varphi= a\wedge b$ 
at step $1$ is $a[1] \wedge b[1]$, where $a[1] \in V_{in}[1]$
and $b[1] \in V_{out}[1]$\@.
We use $\alpha(\varphi)$ for denoting these encodings where $ \alpha={\{a\mapsto a[1],b\mapsto b[1]\}}$.  
More generally, for a variable assignment $\alpha$, 
$ \alpha( \varphi)$ replaces variables $z$ in $\varphi$ 
with $\alpha(z)$\@. 

\begin{figure}[t]
    \begin{align*}
       & A_j(\textbf{G}(\varphi\rightarrow \textbf{X}^i p)) 
            = \theta_1(\omega_j^l), \;\; G_j(\textbf{G}(\varphi\rightarrow \textbf{X}^i p))
        = p[j+i]\\
      &  A_j(\textbf{G}(\varphi\rightarrow \textbf{X}^i (p\textbf{U} \psi)))
            = \theta_1(\omega_j^l)\wedge \neg \theta_2(\omega_{j+i}^l) \\
          &  G_j(\textbf{G}(\varphi\rightarrow \textbf{X}^i (p\textbf{U} \psi)))
        = \bigwedge \limits_{j\leq j' \leq {k-1}}(p[j'+i] \wedge \neg \theta_2(\omega_{j'+i}^l))\wedge    \theta_2(\omega_{k}^l) \\
       & A_j(\textbf{G}(\varphi\leftrightarrow \textbf{X}^i p))
            =  \theta_1(\omega_j^l) \vee p[j+i], \;\;  G_j(\textbf{G}(\varphi\leftrightarrow \textbf{X}^i p))
        = p[j+i]\\            
      & A_j(\textbf{G}p) = \mathit{true},  \;\;  
        G_j(\textbf{G}\phi^0)    
        =  \bigwedge \limits_{n=0}^k   \alpha(\phi^0),
        A_j(\textbf{F}p) = \mathit{true}, \;\; 
    G_j(\textbf{F}\phi^0) 
        = \bigvee \limits_{n=0}^k  \alpha(\phi^0)
    \end{align*}
$\omega_j^l$ is the $l$-length sub-word of the input word for the $l$-length monitor,  $\theta_1$ ($\theta_2$) is used for the word valuation of the monitor for $\varphi$ ($\psi$), and $\alpha={\{v\mapsto v[n]\;|\;,v\in var(\phi^0)\}}$\@.
\caption{Assumption and Guarantee Formulas.}
\label{fig:AG}
\end{figure}

Word evaluations $\theta$ are encoded by means of Boolean  variables. 
In this way, $v_1 \wedge v_2$ and $p\vee q$ are encoded
respectively as the conjunction $v_1[j] \wedge v_2[j]$ and the disjunction $v_1[j] \vee v_2[j]$ 
for each time step $j$, 
and the $i^{\mathit{th}}$-next step  $\textbf{X}^i v_1$ is encoded as $v_1[j+i]$ for $v_1[j]$\@.
This kind of encoding of word valuation is used in the definition of the assumption and guarantee encodings (at step $j$) in Figure~\ref{fig:AG} for a given formula.

Now, for $\mathit{Spec}$ of the form $\bigwedge_{0\leq m'_{E}\leq m}\varrho_{m'_{E}} \rightarrow\bigwedge_{0\leq m'_{S}\leq m'}\eta_{m'_{S}}$, encoding yields 2QBF formulas for $0\leq j \leq K$\@.
\begin{equation}
\begin{aligned}\label{encoding:formula}
\mathit{Cons}^{j, K}(\mathit{Spec})& := 
 \forall V_{in}[j].\,\exists V_{out}[j].\, \\
 &(\bigwedge \limits_{1\leq m'_{E}\leq m}A_j(\varrho_{m'_{E}}) \rightarrow G_j(\varrho_{m'_{E}})) \rightarrow
(\bigwedge \limits_{1\leq m'_{S}\leq m'}A_j(\eta_{m'_{S}}) \rightarrow G_j(\eta_{m'_{S}}))\mbox{\@.}
\end{aligned}    
\end{equation}

 \begin{lemma}[Correctness and Completeness]\label{lemma:completness}
For a GXU specification $\mathit{Spec}$ (in $V = V_{in} \cup V_{out}$) with completeness threshold $K$\@: 
$\mathit{Cons}^{j, K}(\mathit{Spec})$ is satisfiable for all $0\leq j\leq K$ if and only if $\mathit{Spec}$ is realizable.
 \end{lemma}

%

Detailed proofs of the correctness and completeness  of the consistency checks  are included in Appendix~\ref{appendix:lemmasoundness}\@. Consistency checking (\ref{encoding:formula})
is illustrated by the use of the
GXU specification (\ref{ex:assumption}),
which is also used as a running example. 
\begin{example}
\label{example:encoding}
 Let $\varphi_1 :=\textbf{G}(a \rightarrow \textbf{X} b)$, $\varphi_2 :=\textbf{G}(b \rightarrow \textbf{X} \neg b)$ be the given GXU specification with one input variable $a$ and one output variable $b$\@. The completeness threshold for this specification is $K=2$, since the completeness threshold for the individual formulas is $1$ and the length of the corresponding monitor is $2$\@.
 Therefore $\{a[j], a[j+1]\}$ are the universally quantified input variables, and $ \{b[j],b[j+1]\}$ are the existentially quantified output variables, 
  for each $0\leq j \leq 2$\@. 
  From the transition relation of the underlying Mealy machines for
  $\varphi_1$ and $\varphi_2$ we therefore obtain the constraints
 $(a[j] \rightarrow b[j+1] )\wedge  (b[j] \rightarrow  \neg b[j+1])\mbox{\@.}$
As a consequence, the consistency check $\mathit{Cons}^{j,2}( \varphi_1 \wedge \varphi_2)$ in (\ref{encoding:formula}) is obtained as the 2QBF formulas (for $0 \leq j \leq 2$)
\begin{equation}
\begin{aligned}
	\forall  a[j],a[j+1]\,\exists b[j],b[j+1].\,  
	(\neg a[j]\vee b[j+1])  \wedge (\neg b[j]\vee   \neg b[j+1])\mbox{\@.}
\end{aligned}
\label{encodingF}
\end{equation}
The formula in (\ref{encodingF})
is not satisfiable for 
$0\leq j \leq 2$\@. 
Therefore, 
the GXU specification  $\varphi_1 \wedge \varphi_2$ of two mutually dependent formulas is not realizable.
\end{example}



\paragraph{Determining Skolem Functions.}
 A polynomial time consistency check  is based on the notion of {\em candidate Skolem functions}\@. 
 Subsequently, \emph{Skolemization} substitutes existentially quantified output variables in the consistency checks (\ref{encoding:formula})
 with corresponding candidate Skolem function.
 Therefore, Skolemization reduces consistency checks to the validity of universally quantified propositional formulas of the form  $\forall V_{in}[j]. \varphi(V_{in}[j])$\@. 
 For the sake of simplicity, quantifier prefixes are sometimes omitted in the following.
 
 \begin{definition}[Candidate Skolem Function]
	\label{encoding:def.skolem}
 Let  $\mathit{Spec} := E\rightarrow S$  be a GXU specification (in $V = V_{in} \cup V_{out}$)  with completeness threshold $K$\@.
 For output $v \in V_{out}$, define the subset ${S}^v := \{ \psi \in S \,|\, v \in \mathit{vars}(\psi)\}$  of guarantee formulas.
 %
 Now, the {\em candidate Skolem functions} $\mathit{F}(v[j])$ for the existentially quantified output variable $v[j]$ 
 is the conjunction of 
 definitions for $v[j]_e$ in the universal variables in  $\eta_v$ such that $\eta_v\in {S}^v$\@. 
 Based on the variable renaming $\alpha_j :=\{x\mapsto x[j-i] \,|\, x\in var(\varphi) \vee x \in var(\varphi')\} $\@,
 these definitions for $v[j]_e$ with respect to $\eta_v$ are
 obtained as follows\@: 

 \noindent
\begin{minipage}[t]{.5\textwidth}
\begin{equation*}
v[j]_e=
	\begin{cases}
	     \beta & \text{if }j<i\\
		 \alpha_j(\varphi)\vee \beta  & \text{  if } p=v\\
		 \neg \alpha_j(\varphi)\wedge   \beta  & \text{ if } p=\neg v\\
		 \end{cases}
\end{equation*}
\captionof{figure}{$\eta_{v}$ is reaction (\ref{lbl:one-a}) $\textbf{G}(\varphi\rightarrow \textbf{X}^i   p)$ }
\end{minipage}%
\begin{minipage}[t]{.5\textwidth}
	\begin{equation*}
   v[j]_e=
	\begin{cases}
	     \beta & \text{if }j<i\\
		  \alpha_j(\varphi)   & \text{ if } p=   v\\
		 \neg \alpha_j(\varphi)   & \text{ if } p =\neg v\\
		 \end{cases}
	\end{equation*}
	\captionof{figure}{$\eta_{v}$ is invariances (\ref{lbl:three})   $\textbf{G}(\varphi\leftrightarrow \textbf{X}^i   p)$}
\end{minipage}%

\noindent
\begin{minipage}[t]{1\textwidth}
	\begin{equation*}
   v[j]_e=
	\begin{cases}
		 \beta & \text{if } j<i\\
		( \bigvee\limits_{0}^{j}\sigma|_{\alpha,\varphi}  \wedge \neg  \alpha(\varphi'))  \vee   
		\beta &\text{   if } p =v\\
		 \neg (\bigvee\limits_{0}^{j} \alpha(\varphi)  \wedge \neg  \alpha(\varphi')  ) \wedge \beta  & \text{ if   }  p =\neg v\\
	\end{cases}	\end{equation*}
	\captionof{figure}{$\eta_v$ is reaction  (\ref{lbl:one-b})  $\textbf{G}(\varphi\rightarrow \textbf{X}^i   (p\textbf{U} \psi))$}
\end{minipage}\\
\newline
\noindent
Finally,
   $\mathit{CSF}^j(\mathit{Spec}):=\{\mathit{F}(v)\;|\;v\in V_{out}[j],0\leq j\leq K\}$
 collects the set of candidate Skolem functions for all outputs in $\mathit{Spec}$\@.
\end{definition}
Let  $\mathit{Cons}^{j,K}(\mathit{Spec})$ in (\ref{encoding:formula}) be of the form $\forall V_{in}\exists  V_{out}.\varphi( V_{in}, V_{out})$\@.
Then, \emph{Skolemization} yields 
the (implicitly universally quantified) propositional formula
\begin{equation}
	\Phi_{sko}^j(\mathit{Spec}):=  \alpha(\varphi) \text{ where }\alpha=\{ y\mapsto \mathit{F}(y) \,|\, y\in V_{out}[j]\}, \forall 0\leq j\leq K\mbox{\@.}
	\label{propF}
\end{equation}

 


\begin{lemma}
\label{lemmatau}
	 Let $\mathit{Spec}$ be a given GXU specification with completeness threshold $K$.
     The candidate Skolem functions $\mathit{CSF}^j(\mathit{Spec})$ are 
     Skolem functions of the consistency checks 
     $\mathit{Cons}^{j,K}(\mathit{Spec})$ 
     if and only if  
     $\Phi_{sko}^j(\mathit{Spec})$ is valid for all $ 0\leq j \leq K$.
\end{lemma}
Now, the following fact is immediate.
\begin{lemma}
\label{reaLemma}
	 A GXU specification $\mathit{Spec}$ with completeness threshold $K$
  is realizable if and only if  $\mathit{CSF}^j(\mathit{Spec})$ are Skolem functions for the consistency checks $\mathit{Cons}^{j,K}(\mathit{Spec})$, for $0 \leq j \leq K$\@. 
	\end{lemma}
	
	

 \begin{example}\label{ex:skolemization}
 Let's continue with our running Example~\ref{ex:assumption}\@. 
 After skolemization with the  candidate Skolem functions  $b[j]\leftrightarrow a[j-1]$  for formula (\ref{encodingF}), we have $\Phi_{sko}^j(\mathit{Spec}):
 	 (\neg a[j] \vee a[j-1]) 
 	\label{exampleF}
$
 that is not valid, 
 and therefore,  the given specification~(\ref{ex:assumption}) is unrealizable. 
 \end{example} 
\begin{theorem}\label{thm:GXU.polynomial}
Realizability for a GXU specification $E\rightarrow S$ with variables in $V$ is decidable in  $\mathcal{O}(|E|\cdot |S|^2\cdot|V|^3)$\@.
\end{theorem}
\begin{proofsketch}
Validity checking for formula (\ref{propF}) involves verifying the tautology of each clause depending on the number of  variables $V$ and the complexity comes from the cost of traversing a clause i.e. $\mathcal{O}(|V|\cdot (|V|-1))$. For each formula, we have the constraints as the form $A_j \rightarrow G_j$ ($\neg A_j \vee G_j$). As $G_j$ is a conjunction of word evaluations and literals, applying distributive law, we have the number of clauses based on literals and word evaluations that is $\mathcal{O}(|V|)$\@. For a GXU specification $E\rightarrow S$ with variables in $V$, the size of clauses is $(|E|+1)\cdot |S|\cdot |V|\cdot K \cdot |S|$.  So GXU realizability can be decided in $\mathcal{O}(|E|\cdot |S|^2\cdot|V|^3)$\@. A detailed derivation of this complexity bound is listed in 
Appendix~\ref{Appendix:complexity}\@.
\end{proofsketch}

\label{sec:synthesis}
\subsection{Assumption Mining}
In case a consistency check fails, stronger environment assumptions 
are mined for making the synthesis problem realizable based on eliminating the unrealizable core.
\begin{definition}[Unrealizable Core] 
Let $Spec$ be an unrealizable GXU specification over $V_{in} \cup V_{out}$ with a bound $K$ and $\Phi_{sko}^j(Spec)$  (formula (\ref{propF})) be the corresponding propositional formula after Skolemization where $0\leq j \leq K$, an unrealizable core $uc_{\Phi_{sko}}^j$ is   
a CNF formula equivalent to $\Phi_{sko}^j(Spec)$ such that $\forall V_{in}[j].uc_{\Phi_{sko}}^j$ is not valid. 
\end{definition}
%
Based on  Lemma \ref{lemmatau} and Lemma \ref{reaLemma}, one solution to fix the unrealizability is to make each clause in  $uc_{\Phi_{sko}}^j$ valid. 
Let  $\psi$ be a clause  in $uc_{\Phi_{sko}}^j$,
the assumption  should satisfy $\neg \psi'$ to repair unrealizability where $\psi'$ is a 
subformula of $\psi$\@.
%
Using Padoa's  characterization of definability (Lemma~\ref{padoa}), 
one might  check  if one of the variables in $\psi$ is
definable in terms of other variables. 
In these cases, an explicit definition is obtained as the Craig interpolant $I$ of 
  $(\psi \wedge v )\wedge  (\alpha(\psi) \wedge \neg v') $ where $\alpha=\{y\mapsto y'\,|\, y\in var(\psi) \setminus \{v \}\} $\@.
In case every  such variable is "independent",  we follow a different path to extract an assumption. The procedure is illustrated by the following example, along with the concept of the unrealizable core.  
\begin{example}
Let $\Phi_{sko}^j(Spec)=\neg (x_1[j] \wedge x_2[j])\wedge x_3[j]$ be the formula after Skolemization. Then,   $uc_{\Phi_{sko}}^j=\phi_1 \wedge \phi_2$ where $\phi_1=(\neg x_1[j] \vee \neg x_2[j]),\phi_2=x_3[j]$. For $\phi_1$, the formula $\psi= x_1[j] \leftrightarrow \neg x_2[j]$ (variable $x_2$ is  definable in terms of $x_1[j]$)  makes clause $\phi_1$ valid. This can be established by
Padoa's characterization of definability: 
Let  $Z=\{x_2[j]'\}$ be the fresh variable, and we have the following unsatisfiable formula $(\neg (x_1[j] \wedge x_2[j]) \wedge x_2[j]) \wedge (\neg (x_1[j] \wedge x_2[j]') \wedge \neg x_2[j]')$\@.  However, $x_3[j]$ is independent in $\phi_2$. Let $\phi_2=x_3[j]\vee \phi'$  such that $\phi'=\text{false}$. The formula $\psi'=x_3[j] \leftrightarrow \neg \phi'$ is the refinement to make $\phi_2$ valid. Therefore, $\psi \wedge \psi'$ is the  assumption that makes  $\Phi_{sko}^j(Spec)$ valid. 
\end{example}
\begin{algorithm}
  \caption{{Assumption Mining}}\label{refin}
  \begin{algorithmic}[1]
  \Require{Unrealizable core $\bigwedge\limits_{m'=1\cdots m}\varrho_{m'}$}
  \Ensure{A set of refinements $\mathcal{R}$}
  
  \ForAll{$m': 0\leq m'\leq m $}
   $\varphi:=\text{true}$  
  \ForAll{$x\in var(\varrho_{m'})$}
  $\psi \leftarrow \Phi\wedge x$, $\psi' \leftarrow \Phi'\wedge \neg x'$
  \If{isdefined($x,var(\varrho_{m'})\setminus \{x\}$)}
     \State $I  \leftarrow getInterpolant(\psi, \psi')$
     \State $\varphi:=\varphi \wedge (x \leftrightarrow  I  )$
     \State Continue
     \Else{ \If{$\varrho_{m'}$ is of the form $\varphi' \vee \varphi''$}
     \State $\varphi:=\varphi \wedge \{\varphi'' \leftrightarrow \neg \varphi'\}$
     \State Continue
     \EndIf}
  \EndIf
  \EndFor
\If{$\varphi=\text{true}$}
 return $\emptyset$
\EndIf
 \State $\mathcal{R}=\mathcal{R}\cup \{\varphi\}$
  \EndFor
   \State return $\mathcal{R}$
  \end{algorithmic}
\end{algorithm}

The input to the resulting Algorithm \ref{refin}
is an unrealizable core as obtained
from a failed consistency check. 
For such an unrealizable core,  
Algorithm \ref{refin} generates
stronger environment assumptions for making
the given specification realizable.
Algorithm \ref{refin} proceeds by iterating over 
the clauses in the unrealizable core,
and generates assumptions based on applications of Padoa's characterization and Craig interpolation. 
It terminates, since the number of variables is finite. Clearly, if unrealizable cores are minimal then the resulting environment refinements become more precise.
%
We return to our running example, and
continue based on the
developments in the Examples~\ref{example:encoding} and~\ref{ex:skolemization}\@. 
\begin{example}
 $\Phi_{sko}^j(\mathit{Spec})= \neg a[j]\vee a[j-1]$ 
is not valid, and Algorithm~\ref{refin} yields (see Line 9)
the constraint $a[j-1]\leftrightarrow \neg a[j]$\@.
This pointwise constraint might be {\em decoded} as
the GXU formula $\textbf{G}(a\rightarrow \textbf{X}\neg a)$\@. Therefore, adding the assumption $\varphi'=\textbf{G}(a\rightarrow \textbf{X}\neg a)$ to the environment, the specification   $\varphi' \rightarrow \varphi_1 \wedge \varphi_2$ is realizable.  
\end{example}
More generally, the resulting constraints from Algorithm~\ref{refin}
are decoded into the language of GXU formulas as follows.
Let $\psi, \psi_1, \psi_2$ be propositional formulas.
  and $\psi[j]$ be the formula where all the variable $v$ are equipped with time point $j$, that is $v[j]$\@.
  Decoding applies the following rules:
 \begin{enumerate}
 	\item if $x[j]\leftrightarrow\psi[j']$, then: if $\psi\in\{x[j']~|~ j'\neq j \}$, let $\psi'$ be the formula replacing the variables in $\psi$ with the one without the time point, then  $\textbf{G}(x \rightarrow \textbf{X}^{j'-j} \psi')$ is obtained;
\item  if $x[j]\leftrightarrow  (\bigwedge \limits_{j\leq j'\leq k-1}(\neg \psi_1[j'] \wedge \psi_2[j'])) \wedge \psi_1[k]$, then $\textbf{G}(x \rightarrow \psi_2 \,\textbf{U}\,\psi_1)$ is obtained;
\item if $x[j]\leftrightarrow  \bigwedge \limits_{j\leq j'\leq k}\psi[j']$, then $\textbf{G}\psi$ is obtained; 
\item if $x[j]\leftrightarrow  \bigvee \limits_{j\leq j'\leq k}\psi[j']$, then $\textbf{F}\psi$ is obtained. \end{enumerate}
These decoding rules directly correspond to the encodings as discussed above. 
In particular, rule 1 is based on reactions~(\ref{lbl:one-a}), Rule 2 indicates the reaction of the form ~(\ref{lbl:one-b}), Rule 3 is for the global invariance pattern, and Rule 4 is for the liveness pattern~(\ref{lbl:four})\@.

\label{sec:repair}
 \begin{figure}[t]
	 \includegraphics[width=\linewidth]{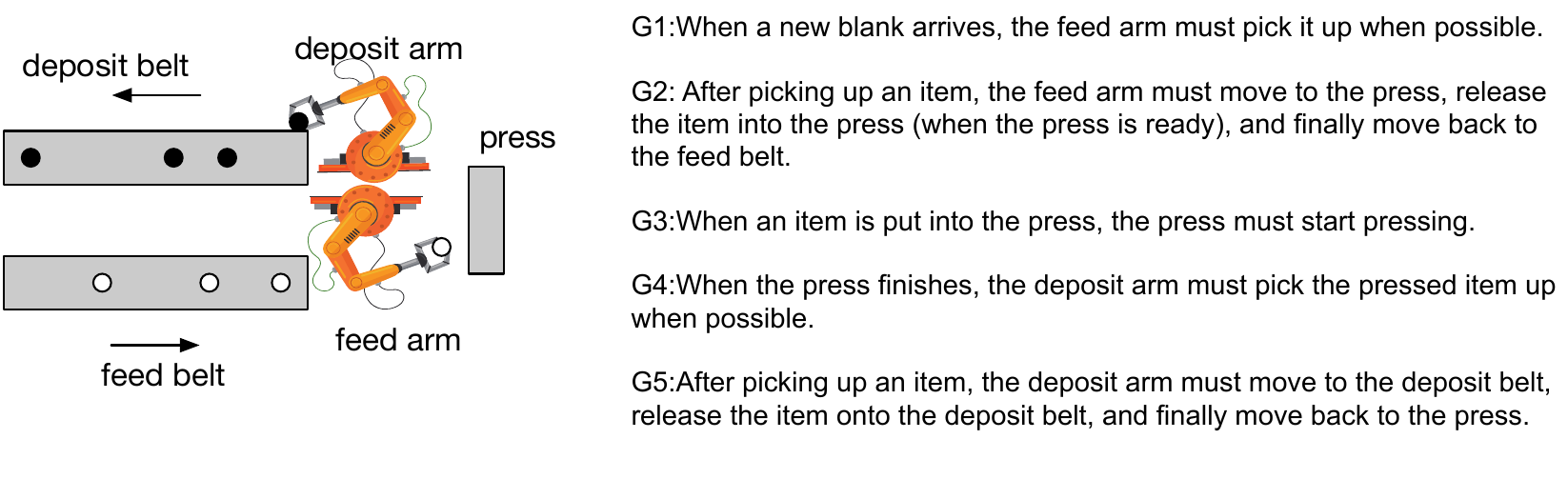}
  \caption{Natural language requirements for production cell~\cite{gritzner2018synthesizing}\@.}
  \label{fig:robot}
\end{figure}

\section{Case Study}
We demonstrate GXU synthesis together with assumption mining using the case study of a  production cell~\cite{gritzner2018synthesizing} (see Figure~\ref{fig:robot})\@. 
The blank workpieces arrive via a feed conveyor equipped with a sensor that signals the arrival of new workpieces to a control system. 
These blanks are then picked up by a robotic arm and placed in a press, which presses the blanks into useful objects. 
The pressed objects are then picked up by another robot arm, which places the objects on a conveyor belt that transports them to the next destination. 
Natural language requirements are summarized in Figure~\ref{fig:robot}. The  following results rely on a combination of 1) autocode4 \cite{cheng2017autocode4}, extended to obtain Mealy machines with monitors, 2) calculator \cite{calculator}, to simplify the propositional formulas for validity checking, and 3) Unique \cite{unique} to definitions.

The exact logical meaning in GXU for  G1-G5 in Figure~\ref{fig:robot} is based on the introduction of a number of input-output variables: (Inputs)
          $blank$, $p.ready$, $i.picked$, $i.release$, $d.loc.deposit$,  $d.loc.press$, $f.loc.feed$, $i.loc.press$, $i.pressed$\@;
      and (Outputs)
           $f.pick$, $f.move$, $d.move$, $d.pick$, $f.release, d.release$, $press$\@.
%
%
A GXU specification for the requirements  is listed as follows\@. 
Hereby, requirement G1 is formalized by the formulas $(\varphi_1)$, $(\varphi_2)$,  
G2 by $(\varphi_3)$-$(\varphi_6)$,  
G3 by $(\varphi_7)$, 
G4 by $(\varphi_8)$, $(\varphi_{9})$, and
G5 by $(\varphi_{10})$-$(\varphi_{13})$\@.
\vspace{-0.08cm}
\footnotesize{
   \begin{align*}
    (\varphi_1)~~  & \textbf{G}(blank \rightarrow \textbf{X}(   \neg f.pick \;\textbf{U}\; f.loc.feed)),\; \; \; \; \; \; \; 
     (\varphi_2)~~ \textbf{G}(f.loc.feed \wedge \neg i.picked \rightarrow \textbf{X}f.pick)\\ 
    (\varphi_3)~~  & \textbf{G}(f.loc.feed \wedge i.picked\rightarrow \textbf{X}( f.move \;\textbf{U}\; (f.loc.press \wedge p.ready))) \\
    (\varphi_4)~~  &\textbf{G}( f.loc.press \rightarrow \textbf{X}(f.move \textbf{U}f.loc.feed)),\; \;
    (\varphi_5)~~ \textbf{G}(f.loc.press \wedge i.picked \rightarrow \textbf{X} \neg f.move)\\
    (\varphi_6)~~  & \textbf{G}(f.loc.press \wedge i.picked \rightarrow \textbf{X}  f.release), \; \;\; \;
     (\varphi_7)~~  \textbf{G}(i.loc.press \rightarrow \textbf{X} press)\\
    (\varphi_8)~~ & \textbf{G}(i.pressed \rightarrow \textbf{X}( \neg d.pick \; \textbf{U}\; d.loc.press)),\;
    (\varphi_{9})~~ \textbf{G}( d.loc.press \wedge \neg i.picked \rightarrow \textbf{X}d.pick)\\
    (\varphi_{10})~~ &\textbf{G}( i.picked \wedge i.pressed  \rightarrow \textbf{X}( d.move \;\textbf{U}\; d.loc.deposit))\\
    (\varphi_{11})~~ & \textbf{G}(d.loc.deposit \wedge i.picked  \rightarrow\textbf{X} \neg d.move),\; \;
    (\varphi_{12})~~   \textbf{G}(d.loc.deposit \wedge i.picked \rightarrow \textbf{X}  d.release)\\
    (\varphi_{13})~~ &\textbf{G}( \neg i.picked \wedge d.loc.deposit \rightarrow \textbf{X}(d.move \;\textbf{U}\; d.loc.press))
  \end{align*}
}
\normalsize

The control program obtained from the GXU specification  is represented by the set of Mealy machines with monitors in Figure~\ref{fig:robot.gxu} (Appendix \ref{appendix:cons})\@. We assume a bound $k$ (similarly, $k_1$, $k_2$, $k_4$ and $k_5$, see below), which  is the completeness threshold for the GXU formula.
Now to check the consistency between ($\varphi_1$)-($\varphi_{13}$),
candidate Skolem functions are generated for all output variables. 
Taking $f.pick$ for example, is:  $\forall m, 1\leq j \leq k$,  
\begin{equation*}
\footnotesize
	 f.pick[j]\leftrightarrow \big((\neg \bigvee\limits_{i=0}^{j-1}blank[i] \vee f.loc.feed[j-1] ) \wedge \beta \big) \wedge \big((f.loc.feed[m-1] \wedge \neg i.picked[j-1])\vee \beta \big) 
\normalsize
\label{case:skolem.formula}	
\end{equation*}
Details of the underlying lengthy calculation can be found in Appendix~\ref{appendix:cons}\@.
Skolemization results in a validity problem.
For the GXU formula $\varphi_1$, for example, this step leads to the formula
\begin{equation}
\footnotesize
 \begin{aligned}
		 	&    (\neg blank[j] \vee f.loc.feed[j+1]) \vee \big( \bigwedge_{j'=j}^{ k_{1}-2} \Big( \neg f.loc.feed[j'+1] \wedge \neg \big((\neg (\bigvee\limits_{i=0}^{j}(blank[i]  \\
		 	&\vee f.loc.feed[i+1])) \wedge \neg i.pick[j'] \wedge f.loc.feed[j'+1] \wedge \beta)\big) \big)\\
		 	\end{aligned}
		 	\label{case:prop.formula}
		 	\end{equation} 

The resulting consistency check (see Appendix \ref{appendix:unsat}), however, fails and yields
an unrealizable core (again, see Appendix \ref{appendix:unsat})\@. The unrealizable core corresponding to the formula $\varphi_{10}$ is the CNF formula:
    \begin{equation}
    \footnotesize
    	\begin{aligned}
    		&   (\bigwedge_{j'=j}^{ k_{1}-2}(\neg blank[j] \vee f.loc.feed[j+1] \vee 
    		 \neg f.loc.feed[j'+1] \vee i.pick[j'] ))\\
    		 &   
	\wedge (\neg blank[j] \vee f.loc.feed[j+1] \vee  f.loc.feed[k_1])
    	\end{aligned}
    	\label{case:unsat.formula}
    \end{equation}
Applying Padoa's characterization of definability (line 3 of Algorithm 1) to the unrealizable core ($\ref{case:unsat.formula}$) for $blank[j]$, 
one obtains satisfiable result,
that is, values for $blank[j]$ can be assigned, in some sense, independently.
Lines 8-11 of the repair Algorithm~1 are then applied to obtain:
   \begin{align}
   \label{case:refinement1} & blank[j]\leftrightarrow f.loc.feed[j+1]\\
  \label{case:refinement2}  &  blank[j]\leftrightarrow (\bigwedge_{j'=j}^{ k_{1}-2}\neg f.loc.feed[j'+1] \wedge \neg i.picked[j']) \wedge  f.loc.feed[k_1])
    \end{align}
Now, formula (\ref{case:refinement1}) is discarded as it results in a vacuous refinement, but decoding formula (\ref{case:refinement2}) yields a sensible
temporal constraint on inputs. 
\begin{align}\label{env.cnstrnt}
	\textbf{G}(blank &\rightarrow \neg i.picked\;\textbf{U}\;f.loc.feed)\mbox{\@.}
\end{align}
Processing the remaining unrealizable clauses in Algorithm~1 leads to three further  validity problems.
     \begin{align*}
    &  i.pressed[j]  \leftrightarrow (\bigwedge_{j'=j}^{ k_{4}-1}\neg i.picked[j'] \wedge d.loc.press[j']) \wedge d.loc.press[k_4] )\\
     & i.picked[j]\wedge i.pressed[j]\leftrightarrow   \bigwedge_{j'=j}^{ k_{5}-1}(\neg d.loc.deposit[j'] \wedge \neg d.loc.press[j']))  \wedge d.loc.deposit[k_5] )\\
     &  i.picked[j]\wedge i.pressed[j]\leftrightarrow   \bigwedge_{j'=j}^{ k_{5}-1} \neg i.picked[j']
     \end{align*}
Now, the decoding step for these formulas in Algorithm~1 leads to three additional environment constraints.
  	\begin{align}
     \label{env.cnstrnt2} \textbf{G}(i.pressed &\rightarrow   \neg i.picked \; \textbf{U } d.loc.press)  \\
     \label{env.cnstrnt3} \textbf{G}(i.picked \wedge i.pressed &\rightarrow   \neg d.loc.press \; \textbf{U } d.loc.deposit) \\
     \label{env.cnstrnt4}\textbf{G}(i.picked \wedge i.pressed &\rightarrow \text{ true}\; \textbf{U } \neg i.picked)   
     \end{align}
Altogether, constraining the control environment of the production cell in Figure~\ref{fig:robot.gxu} with the mined assumptions
 (\ref{env.cnstrnt}), (\ref{env.cnstrnt2}), (\ref{env.cnstrnt3}), and (\ref{env.cnstrnt4}) 
 yields a realizable specification\@.

\label{sec:case.study}
\section{Related Work}
GXU synthesis is strongly based on the {\em structural} 
approach to reactive synthesis~\cite{CYR16,cheng2017autocode4}\@. 
But GXU synthesis supports an expressive rule-like logic including {\em strong until}, 
it is complete for mutually dependent specifications, and it is
decidable in polynomial time in the size of GXU specifications (Theorem~\ref{thm:GXU.polynomial})\@.
Real-time extensions and other pragmatics for synthesizing programmable logic controllers~\cite{xie2022structural} can be adapted to the more expressive GXU logic.

Structural synthesis~\cite{CYR16,cheng2017autocode4} is based on a reduction of the realizability to the 2QBF validity problem, which is $\Pi_2^p$-complete.
A more detailed analysis of the resulting 2QBF 
realizability conditions with the aim of obtaining improved
complexity bounds for these kinds of subclasses has been missing.
Reactive synthesis based on solving 2QBF 
has also been considered in~\cite{katis2018validity,katis2020formal},
which is based on greatest fixpoints to represent acceptable control behavior. Again, these logic-based approaches to reactive synthesis rely on general
2QBF solving.

Since GXU contains GR(1) as a sublogic, the polynomially bounded runtimes of GXU synthesis in Theorem~\ref{thm:GXU.polynomial} improve the known complexity result for GR(1) synthesis~\cite{piterman2006synthesis}\@, 
which, even if it is cubic in an underlying state space, may well be non-polynomial with respect to the problem statement itself. 
This observation also holds for other LTL fragments~\cite{asarin1998controller}\@.


Finite LTL synthesis~\cite{kupferman2001model,de2015synthesis,xiao2021fly,zhu2017symbolic} generates deterministic finite automata (DFA)  for $\text{LTL}_f$ specifications, that is, LTL formulas interpreted over finite traces.
The $\text{LTL}_f$ to DFA translation in~\cite{kupferman2001model} is doubly-exponential in the size of the formula, 
\cite{de2015synthesis} reduces $\text{LTL}_f$  to DFA synthesis
to reachability games, which are 2EXPTIME complete,
optimizations for $\text{LTL}_f$ synthesis have been studied in \cite{zhu2017symbolic},
and $\text{LTL}_f$ satisfiability checking  is applied as a preprocessing step~\cite{li2019sat}, thereby obtaining potentially exponential speedups~\cite{xiao2021fly}\@.
There are two main differences between GXU and $\text{LTL}_f$ synthesis\@. 
First, GXU synthesis is designed to work only for a syntactically restrictive {\em assumption-guarantee} fragment of LTL formulas, 
while the $\text{LTL}_f$ synthesis supports all LTL formulas, albeit in a finitary interpretation.
Second, a polynomial time GXU synthesis algorithm is obtained from these restrictions, whereas $\text{LTL}_f$ synthesis is doubly-exponential. 
It might be interesting, however,
to possibly extend the core GXU synthesis algorithm to work for all $\text{LTL}_f$ formulas, or, alternatively, to recover a polynomial GXU-like synthesis algorithm for certain fragments in the general framework of $\text{LTL}_f$ synthesis\@. 


Minimal unrealizable cores have previously been used 
to locate errors in an unrealizable specification, based
on approximations of the winning region by a minimization algorithm~\cite{konighofer2010debugging}\@.
In contrast, our unrealizable core are purely logical, and
logical operations are used to synthesize repairs to unrealizability.
 Counterstrategies for synthesis games are generally used to exclude unrealizable environmental
behavior~\cite{li2011mining,cheng2014g4ltl,chatterjee2008environment,alur2015pattern,alur2013counter,ehlers2012symbolic}\@.
In this way, predefined templates were proposed, which are instantiated on the basis of the exclusion of counter-strategies~\cite{li2011mining,cheng2014g4ltl}\@.
Moreover, a 
counterstrategy-driven refinement loop based 
on explicit value abstraction, Sa{\"i}di-Graf predicate abstraction, and Craig interpolation for mining assumptions
is described in~\cite{hajdu2016configurable}\@.
In contrast to the algorithm for mining assumptions in Section~\ref{sec:repair}, these approaches rely on suitable expert input, such as specification templates and also predefined scenarios.
Minimal assumptions have been 
proposed for repairing unrealizable specifications~\cite{chatterjee2008environment,alur2015pattern,alur2013counter}\@. 
But the computation of minimal liveness assumptions is NP-hard.
Assumption mining as developed in this paper has been influenced by 
recent work on constructing explicit definitions in the context of preprocessing and solving QBF formulas~\cite{slivovsky2020interpolation}.

\label{sec:related}
\section{Conclusions}

Structural synthesis for GXU is attractive for autogenerating correct-by-construction control programs. 
First, it supports the rule-like control specifications that are often found in stylized requirement specification languages for embedded control (EARS~\cite{Mav19}, CLEAR~\cite{hall2018clear}, FRET~\cite{dimitrova2019synthesizing}) and also for information 
systems~(Rimay~\cite{veizaga2021systematically})\@.
Second, generated programs may be traced back
to specifications as mandated by applicable
safety engineering codes (IEC 61508, DO178C, ISO 26262)\@.
Third, the generated reactive programs are synchronous, and can therefore readily be integrated into current design flows by further compiling them into widely used design languages such as Scade, LabView, Simulink, and continuous function charts (IEC 61131-3)~\cite{cheng2017autocode4,cheng2014g4ltl}\@.
Alternatively, they might also be compiled into 
performant concurrent programs~\cite{lee1987synchronous,lohstroh2021toward}\@.

Control problems in the real world usually depend on real-time constraints.
These can easily be added to GXU synthesis by considering Mealy machines extended with event clock constraints, and by consistency checks based on satisfiability modulo, say, 
an efficiently decidable theory of clock constraints.
Similarly, the support of {\em data types} such as linear and non-linear arithmetic is essential for the specification of many {\em hybrid} and {\em resilient} controllers.
But now solvers for a larger class of domain constraints can dominate the complexity of the underlying core synthesis algorithm.

The mining of assumptions on the basis of 
unrealizable 
cores has the potential to identify 
root causes of unrealizability.
But we need more experience for capturing intended specifications 
%
based on both {\em strengthening} environment assumptions and, dually, {\em weakening}  
control guarantees.
\label{sec:conclusion}
\bibliographystyle{unsrt}
\bibliography{refinement.bib}
\clearpage
\appendix
\chapter*{Appendix}




\section{Proofs}
\subsection{Proof of Lemma \ref{lemma:completness}}
\label{appendix:lemmasoundness}
\textbf{Lemma \ref{lemma:completness} (Correctness and Completeness)}
For a GXU specification $\mathit{Spec}$ (in $V = V_{in} \cup V_{out}$) with completeness threshold $k$\@: 
$\mathit{Cons}^{j, k}(\mathit{Spec})$ is satisfiable for all $0\leq j\leq k$ if and only if $\mathit{Spec}$ is realizable.
\begin{proof}
For each $j$,  let $A_j(\varrho_{k'_{E}}) \rightarrow G_j(\varrho_{k'_{E}})$ be  the constraints of a Mealy machine obtained by the environment and $A_j(\varrho_{k'_{S}}) \rightarrow G_j(\varrho_{k'_{S}})$ be the one obtained by the system. 
 
     $\Rightarrow:$ we proceed the proof that if for every  $0 \leq j \leq k$, $w_j$ over input variables and  $w_j'$ over output variables are satisfying $Cons^{j,k}$, then $q$ is one of the accepting states where $(q,\langle w_j,w_j' \rangle)$ is a configuration represented by the Mealy machine with monitors. For simplifying, we use $A_j \rightarrow G_j$ to present the general case. By construction mentioned in Section \ref{sec:realization}, for each time point, the Mealy machine in Figure \ref{fig:mealymachine}(2) and Figure \ref{fig:mealymachine}(3) present such constraints: if $A_j$ doesn't hold, the Mealy machines stay in $q_0$, otherwise, it moves from $q_0$ to $q_1$ that is $A_j\wedge G_j$. Thus, we have that for each Mealy machines  in Figure \ref{fig:mealymachine}(2) and Figure \ref{fig:mealymachine}(3) there must exist a configuration $(q,\langle w_j,w_j' \rangle)$ presented where  $q\in \{q_0,q_1\}$ which are both accepting states. For the reaction pattern 1b, the corresponding Mealy machine is $\mathcal{T}_1||\mathcal{T}_2$ where $\mathcal{T}_1$ is the similar to the previous case. $G_j$ contains constraint $\theta_2(w_k^l)$ that ensures $T_2$ moves to $q_0$ or $q_2$. So each Mealy machine has such a configuration as $(q,\langle w_j,w_j' \rangle)$. Since $Cons^{j,k}$ is satisfiable, the consistency within these formulas is ensured i.e. no conflict within these configurations. The specification $Spec$ is realizable.

$\Leftarrow:$ we prove that if every accepting configuration $(q,\langle w_j,w_j'\rangle)$ such that $c_0\Rightarrow_{\mathcal{T}}^*(q,\langle w_j,w_j'\rangle)$ for every Mealy machine exist without any conflict i.e. the $Spec$ is realizable, then $Cons^{j,k}$ is satisfiable.  By construction, $ A_j \rightarrow G_j$ can be directly extracted by traversing the Mealy machine in Figure \ref{fig:mealymachine}. That is, $ A_j \rightarrow G_j$ holds for every $j$ and there is no conflict i.e.  $Cons^{j,k}$ is not unsatisfiable.  The statement holds.
\end{proof}
\subsection{Proof of Lemma \ref{lemmatau}}
\textbf{Lemma \ref{lemmatau}}  
 Let $\mathit{Spec}$ be a given GXU specification with completeness threshold $k$.
     The candidate Skolem functions $\mathit{CSF}^j(\mathit{Spec})$ are 
     Skolem functions of the consistency checks 
     $\mathit{Cons}^{j,k}(\mathit{Spec})$ 
     if and only if the propositional formula 
     $\Phi_{sko}^j(\mathit{Spec})$ is valid for all $ 0\leq j \leq k$.
\begin{proof}
	$\Rightarrow:$  We proceed the proof that if the candidate Skolem functions $CSF^j(Spec)$ are Skolem functions of the consistency checks $Cons^{j,k}$, the propositional formula \(\Phi_{sko}^j\) is valid. Assume that the candidate Skolem functions are real Skolem functions i.e. for every existential variable $y$, the candidate Skolem function $F(y)$ is the Skolem function. Let \(\exists V_{out}. \varphi(V_{in}, V_{out})\) be an existential quantifier in \(\forall V_{in}\exists V_{out}.\varphi(V_{in}, V_{out})\), and let \(\mathit{CSF}^j(\mathit{Spec})\) be its corresponding candidate Skolem function.
     By the validity of the candidate Skolem functions, \(\mathit{CSF}^j(\mathit{Spec})\) satisfies \(\varphi(V_{in},F(y))\) for all \(V_{in}, F(y)\in\mathit{CSF}^j(\mathit{Spec})\). Therefore, \(\exists V_{out}. \varphi(V_{in}, V_{out})\) is true for all assignments of universal variables, making \(\Phi_{sko}^j\) a tautology. Thus,  the statement holds from this direction.

	$\Leftarrow: $ We proceed the proof that if \(\Phi_{sko}^j(Spec)\) is valid, the candidate Skolem functions $CSF^j(Spec)$ are Skolem functions of the consistency checks $Cons^{j,k}$.  Assume that \(\Phi_{sko}^j(Spec)\) is valid i.e. the propositional formula \(\Phi_{sko}^j(Spec)\) is a tautology.  Let \(\exists V_{out}. \varphi(V_{in}, V_{out})\) be an existential quantifier in \(\forall V_{in}\exists V_{out}.\varphi(V_{in}, V_{out})\), and let \(\text{CSP}^j(Spec)\) be its corresponding candidate Skolem function. Since \(\Phi_{sko}^j(Spec)\) is a tautology, for any assignment of values to \(V_{in}\), \(\exists V_{out}. \varphi(V_{in}, V_{out})\) is true. This implies that \(\mathit{CSF}^j(\mathit{Spec})\) satisfies \(\varphi(V_{in}, F(y))\) for all \(V_{in}, F(y)\in \mathit{CSF}^j(\mathit{Spec})\). Therefore, the candidate Skolem functions are valid solutions for the existential quantifiers, as they make the formula true for all assignments of universal variables. Thus, $CSF^j(Spec)$ are Skolem functions of the consistency checks $Cons^{j,k}$.
     
     Thus, Lemma \ref{lemmatau} holds.
\end{proof}
\subsection{Proof of Lemma \ref{reaLemma}}
\textbf{Lemma \ref{reaLemma}}
 A GXU specification $\mathit{Spec}$ with completeness threshold $k$
  is realizable if and only if  $\mathit{CSF}(\mathit{Spec})$ are Skolem functions for the consistency checks $\mathit{Cons}^{j,k}(\mathit{Spec})$, for $0 \leq j \leq k$\@.
\begin{proof}
	$\Rightarrow:$ Assume $E\rightarrow S$ is realizable i.e. there exists a set of Mealy machines with monitors realizing  $E\rightarrow S$ and there is no conflict within variables. Thus, the 2QBF formula $Cons^{j,k}=\forall V_{in}\exists  V_{out}.\varphi( V_{in}, V_{out})$ is valid i.e.  for any assignment of values to $V_{in}$, the existential quantifier is satisfied. Let $\exists V_{out}. \varphi(V_{in}, V_{out})$ be an existential quantifier in $\mathit{Cons}^{j,k}(\mathit{Spec})$, and let $\mathit{CSF}^j(\mathit{Spec})$ be its corresponding candidate Skolem function. So we have $F(y)\in \mathit{CSF}^j(\mathit{Spec})$ acts as a witness for $\varphi(V_{in}, V_{out})$.
    Therefore, $\mathit{CSF}^j(\mathit{Spec})$ are Skolem functions for $\Psi$ under which the specification is realizable.

    $\Leftarrow:$ Assume $S_y$ are the Skolem functions.
    Let $\exists V_{out}. \varphi(V_{in}, V_{out})$ be an existential quantifier in $\mathit{Cons}^{j,k}(\mathit{Spec})$, and let $\mathit{CSF}^j(\mathit{Spec})$ be the corresponding Skolem function.
    Since $F(y)\in \mathit{CSF}^j(\mathit{Spec})$ are Skolem functions, they satisfy the existential quantifiers in $\mathit{Cons}^{j,k}(\mathit{Spec})$, and $F(y)$ acts as a witness for $\varphi(V_{in}, V_{out})$ for all $V_{in}$.
    Therefore, there exists a strategy such that for any assignment of values to $V_{in}$, the existential quantifier is satisfied. So we have the 2QBF $Cons^{j,k}$ is valid i.e. there is no conflict within the Mealy machines. $E\rightarrow S$ is realizable.
    
    Thus, Lemma \ref{reaLemma} holds.
\end{proof}

\subsection{Proof of Theorem 1}
\label{Appendix:complexity}
Checking the validity of propositional formula (\ref{propF}) involves verifying the tautology of each clause when it is in CNF. This process is accomplished by examining every literal $x$ in each clause to determine if a corresponding negation literal,  $\neg x$, exists. The size of literal $x$ in one clause depends on the number of all the variables i.e. $|V|$ such that $V=\{v\;| v\in (V_{in}[j]\cup V_{out}[j]),0\leq j \leq k  \}$. Thus, the complexity comes from the cost of traversing a clause which can be finished in time $\mathcal{O}(|V|\cdot (|V|-1))$ and the number of clauses we would have. For one formula, we have the constraints as the form $A_j \rightarrow G_j$, i.e. $\neg A_j \vee G_j$. Because $G_j$ is a conjunction of word evaluations and literals, applying distributive law, we have the number of clauses based on the size of literals and word evaluations that is $\mathcal{O}(|V|)$\@. For a GXU specification $E\rightarrow S$ with variables in $V_{in}\cup V_{out}$, the size of clauses   is $(|E|+1)\cdot |S|\cdot |V_{in}\cup V_{out}|\cdot k \cdot |S|$. The cost is  $(|E|+1)\cdot |S|\cdot |V_{in}\cup V_{out}|\cdot k \cdot |S| \cdot  (|V_{in}\cup V_{out}|\cdot k)^2$. The $k$ is small enough. So the tautology checking can be decided in $\mathcal{O}(|E|\cdot |S|^2\cdot|V_{in}\cup V_{out}|^3)$.  The detailed complexity argument is provided as follows:

	  First, we review the procedure: For a given GXU formula $\varphi_{E}\rightarrow \varphi_{S}$ with input variable $V_{in}$ and $V_{out}$. 
	\begin{enumerate}
		\item To construct a Mealy machine with monitors. For the monitors, it will be $i+1$ step i.e. $\mathcal{O}(i+1)$. Then the procedure to build the Mealy machine 4 steps. Since the complexity to construct a Mealy machine is $\mathcal{O}(2^m)$ where $m$ is the number of the states. In our case, m is 2. It cost $4(i+1)$ to get a Mealy machine with the monitors.
		\item The second step is to the threshold $k$ of the GXU specification.
		\item tautology checking: determining skolem function can be reduced to tautology checking in our case. It can be solved as follows: Let $\Phi$ be in CNF format i.e. $\bigwedge\limits_{m=1\cdots k}\varrho_m$ \begin{enumerate}
   \item For each clause $\varrho_m=l_{m,1}\vee l_{m,2} \cdots l_{m,n}$ being the disjunction of literals.
	\item If there are some $n_1,n_2$ s.t. $l_{m,n_1}=\neg l_{m,n_2}$, mark the clause as a tautology
	\item If all clauses are marked, tautology, otherwise, not tautology.
\end{enumerate} 
		\item Then we need to check the size of encoded formula from the Mealy machine. The number of existential variable $V_{out}\times k$ and the size of universal variable is $V_{in}\times k$. For the further computation of the complexity, we first compute the size of the formula in CNF format.  For every formula, we have:
		\begin{itemize}
			\item  reaction pattern (a):the size of clause of the formula is depending on the number of clause of $(output[j'+1]\vee \neg m_1(\omega_{j'}))$. Since it is well defined, the number of clauses depends on the number of $\wedge$ (size of the candidate Skolem function) i.e. the number of formulae corresponding to the output variable that is $|S|$ and the number of $\wedge$ in each case of the candidate Skolem function whose worst case is for the strong until pattern with $|V_{in}\times k|$. So we have worst case $(|V_{in}\times k|)\times |S|$.  
			The cost of tautology checking for each clause is $N\times (N-1)$ where $N$ is the number of literals for each clause. The worst case of $N$ is $(|V_{in}|+|V_{out}|)\times k$. So we have the cost $\big((|V_{in}\times k|)\times |S|\big) \times \big((|V_{in}|+|V_{out}|)\times k\big) \times ((|V_{in}|+|V_{out}|)\times k)-1) $.
			\item  reaction pattern (b): similar to the previous case. The constraints is for the form $\neg \theta_1(\omega_j)\vee   \theta_2(\omega_{j+i}) \vee (\bigwedge_{j\leq j' \leq {k-1}}(\text{output}[j'+i] \wedge \neg \theta_2(\omega_{j'+i}))\wedge    \theta_2(\omega_{k}) )$. We don't need to unroll the formula. It can be checked by the induction of the $(\text{output}[j'+i] \wedge \neg \theta_2(\omega_{j'+i}))$ for $j'<k$. As mentioned above,   the number of clauses for the candidate Skolem function is $(|V_{in}\times k|)\times |S|$. Taking $\neg \wedge_2(\omega)$ into consideration, the worst case for $\wedge$ will be $(|V_{in}\times k|)$. So the worst number of clauses is $(|V_{in}\times k|)+(|V_{in}\times k|)\times |S|$. The cost of tautology checking for each clause is $N\times (N-1)$ where $N$ is the number of literals for each clause. The worst case of $N$ is $(|V_{in}|+|V_{out}|)\times k$. Then we have $((|V_{in}|+|V_{out}|)\times k)\times (|V_{in}|+|V_{out}|)\times k-1) \times (|V_{in}\times k|)+(|V_{in}\times k|)\times |S|)$.
			\item The liveness pattern only costs $(|V_{in}\times k|)+(|V_{in}\times k|)\times |S|$.
		\end{itemize}
		
For a GXU specification $E\rightarrow S$ i.e. $\wedge_{1\leq j\leq m}\varrho_j \rightarrow \wedge_{1\leq j'\leq n}\eta_j'$, the size of clauses for $\vee_{1\leq j\leq m} \neg \varrho_j \vee \wedge_{1\leq j'\leq n}\eta_j'$ will be $(m+1)\times n \times ((|V_{in}|+|V_{out}|)\times k)\times (|V_{in}|+|V_{out}|)\times k-1) \times (|V_{in}\times k|)+(|V_{in}\times k|)\times |S|)$. So, for the worst case, we have $(|E|\times|S|^2)\times((|V_{in}|+|V_{out}|)\times k)\times (|V_{in}|+|V_{out}|)\times k-1) \times (|V_{in}\times k|)+(|V_{in}\times k|))$. 
 
 Then, we check the computation of the bound $k$. 
 The cost of traversing a clique is shown in Table 1 where $d$ is the diameter, $rd$ is the recurrence diameter and $n$ is the number of accepting states. Thus, the bound for reactive pattern (a) is $2*3$ i.e. 6. 

	\end{enumerate}
	
	Thus, the consistency checking can be solved in  $\mathcal{O}(|E|\cdot |S|^2\cdot|V_{in}\cup V_{out}|^3)$.
 
 \subsubsection{Linear Completeness Threshold Computation}
 \label{appendix:completeness}
 The completeness threshold for the Mealy machines in Figure \ref{fig:mealymachine} is based on the over-approximating the cost of traversing a clique proposed in \cite{lct} where $n$ is the number of states, $d$ is the largest distance between any two reachable states and $rd$ is  the length of a longest simple (loop-free) path.  In our case, the clique containing an accepting state is both vacuously labelled and accepting.  
 
 \begin{table}
 \label{cost:bound}
\centering
    \begin{tabular}{|c|c|c|c|}
       \hline
        C vacuously labelled? & C accepting & cost[C] & $\text{cost}_f[C]$ \\
        \hline
        no & no & $rd+1$&$\infty$ \\
        no & yes & $rd+1$&$(n+1)(rd+1)$ \\
        yes & no & $d$&$\infty$ \\
        yes & yes & $d$&$(n+1)d$ \\
        \hline
    \end{tabular}
    \caption{Over-approximating the cost of traversing a clique \cite{lct}}
\end{table}

Taking the GXU formula shown in Section 4 in a reaction pattern for example (Figure. \ref{fig:fmealymachine}). The given formula is $\textbf{G}(a \wedge \textbf{X}b \rightarrow \textbf{X} b)$. The number of states is 2. they are vacuously labelled and accepting. The $d$ in Table 1 is $1$. According to Table 1, we have the completeness threshold for Mealy machine in Figure \ref{fig:fmealymachine}(A) is $1$. Accordingly, the completeness threshold for the specification is 2 as the length of the monitor is 2. 

Taking the GXU formula in reaction pattern \ref{lbl:one-b} for example. The Mealy machines are in Figure \ref{fig:mealymachine}(1) and  Figure \ref{fig:mealymachine}(4). The completeness of the one in Figure \ref{fig:mealymachine}(1) is the same to the example before. The one for Figure \ref{fig:mealymachine}(4) involves a non-accepting state. Thus, the cost of visiting final clique is $(n+1)d$ where the number of states $n$ is $3$, $d$ is $2$ and the completeness threshold is $6$.

\section{Example of Lift Controller}
\begin{figure}[t]
\begin{align*}  
  (\phi_{1})~~  &\textbf{G}(out_1 \vee out_2) 
\\ (\phi_{2})~~  &\textbf{G} ((b_1\wedge f_1)\rightarrow \textbf{X}(out_1 \textbf{U} \neg b_1 \wedge  f_{2})) \\
  (\phi_{3})~~  &\textbf{G} ((b_2\wedge f_2)\rightarrow \textbf{X}(out_1 \textbf{U} \neg b_2 \wedge (f_{1}\vee f_{3}))
  \\ (\phi_{4})~~  &\textbf{G} ((b_3\wedge f_3)\rightarrow \textbf{X}(out_1 \textbf{U} \neg b_3 \wedge f_{2})\\
  (\phi_{5})~~  &\textbf{G} ((b_1\wedge f_{2})\rightarrow \textbf{X}(out_1 \textbf{U} f_1)) 
 \\(\phi_{6})~~  &\textbf{G} ((b_2\wedge (f_{1} \vee f_{3}))\rightarrow \textbf{X}(out_1 \textbf{U} f_2))\\
  (\phi_{7})~~  &\textbf{G} ((b_3\wedge f_{2} \rightarrow \textbf{X}(out_1 \textbf{U} f_3)) 
 \\ (\phi_8)~~   &\textbf{G} (\neg b_1 \vee \neg b_2 \vee \neg b_3 \rightarrow out_2)
\end{align*}
\caption{Reactive GXU specification for lift controller.}
\label{fig:list}
\end{figure}
\begin{example}[Lift Controller~\cite{cavezza2017interpolation}] 
A lift moves between three floors. 
The lift can only move one floor in each time step. For simplicity, we do not consider the case where the lift is on the third floor and button 1 is pressed, or the lift is on the first floor and button 3 is pressed.
The input variables are $b_i$, for $i=1,2,3$, for representing the status of the
three buttons on each floor, and the variables $f_i$, for $i=1,2,3$, 
are capturing the floor position of the lift. 
The output variables of the controller are  $out_1$ and $out_2$,  
where $out_1$ indicates that the motor is turned on, whereas $out_2$ denotes that the motor is turned off. 
The resulting GXU specification in Figure~\ref{fig:list} does not contain any environment assumptions\@. 
Together, the formulas $\phi_2, \phi_4$ and $\phi_7$ can be read as follows:
if the elevator is on the first or third floor and the button of the same floor is pressed then the motor is switched on until the elevator reaches the second floor.
\end{example}
\newpage 
\section{Details of Case study}
\label{appendix:cons}

The input output is as follows: input $blank$ holds if a new blank arrives, $f.loc.feed$ holds if the feed arm is at the feed belt, $i.pick$ holds if the item can be picked up, $i.picked$ holds if the item is picked up, $i.loc.press$ holds if the item is at the press,
$d.loc.deposit$ holds if the deposit arm is at the deposit belt, and 
$f.release$ holds if the feed arm releases an item to the press. Output $f.pick$ holds if the feed arm picks an item up, $d.move$ holds if the deposit is moving, $f.release$ holds if the feed arm releases an item and $press$ holds if the press is working.
The propositional part of 2QBF can be obtained as:

	\begin{align*}
	\varphi_1:	&\big(( blank[j]\wedge \neg f.loc.press[j+1])  \rightarrow  \bigwedge_{j\leq j'\leq k\-1}  ((f\_arm\_move[j'+1]\\&\wedge \neg f.loc.press[j'+1]) \wedge f.loc.press[k']\\
	 \varphi_2:&( f.loc.feed[j] \wedge \neg i.picked[j]  \rightarrow  f\_arm\_pick[j+1])\\
	\varphi_3:&\big(( f.loc.feed[j]\wedge i.picked[j]) \wedge \neg (f.loc.press[j+1]\wedge press\_ready[j+1])   \rightarrow \\ & \bigwedge_{j\leq j'\leq k'-1}  ((f\_arm\_move[j'+1]  \wedge \neg (f.loc.press[j']\wedge press\_ready[j']) \\ &\wedge (f.loc.press[k']\wedge press\_ready[k']) \big)\\
	\varphi_4:&\big((  f.loc.press[j] \wedge \neg f.loc.feed[j]) \rightarrow   \bigwedge_{j\leq j'\leq k'-1} f.move[j'+1]\wedge \neg f.loc.feed[j'+1]) \\
	\varphi_5:&\big(( f.loc.press[j] \wedge i.picked[j]  \rightarrow  \neg f.move[j+1]\big)\\
 \varphi_6:&\big(( f.loc.press[j] \wedge i.picked[j]  \rightarrow  f\_arm\_release[j+1]\big)\\
	\varphi_7:	& \big((  \neg i.picked[j]\wedge f.loc.press[j] \wedge \neg f.loc.feed[j]) \rightarrow   \\ &\bigwedge_{j\leq j'\leq k' \leq k} f.move[j'+1]\wedge \neg f.loc.feed[j']) 
			\wedge  \bigvee_{i\leq i'\leq k} f.loc.feed[i']\\
\varphi_8:	&\big(( i.loc.pressed[j]  \rightarrow  press[j+1]\big)\\
	\varphi_9:	& \big((i.loc.pressed[j]\wedge \neg d.loc.press[j] ) \rightarrow   \\&\bigwedge_{j\leq j'\leq k' \leq k}  \neg d\_arm\_pick[j'+1]\wedge \neg d.loc.press[j']) 
			\wedge  \bigvee_{i\leq i'\leq k} d.loc.press[i']\\
\varphi_{10}:	 &\big(( d.loc.press[j] \wedge \neg i.picked[j] \rightarrow  d\_arm\_pick[j+1]\big)\\
 \varphi_{11}:		& \big((i.picked[j] \wedge i.loc.pressed[j]\wedge \neg d.loc.deposit[j] ) \rightarrow   \\ &\bigwedge_{j\leq j'\leq k' \leq k} d.move[j'+1] \wedge \neg d.loc.deposit[j']) 
		\wedge  \bigvee_{i\leq i'\leq k} d.loc.deposit[i']\\
  \varphi_{12}:	 &\big(( d.loc.deposit[j] \wedge  i.picked[j] \rightarrow  \neg d.move[j+1]\big)\\
	 \varphi_{13}:& \big(( d.loc.deposit[j] \wedge  i.picked[j] \rightarrow   d\_arm\_release[j+1]\big)\\
	\varphi_{14}:& \big((\neg i.picked[j] \wedge d.loc.deposit[j]\wedge \neg d.loc.press[j] ) \rightarrow \\ &  \bigwedge_{j\leq j'\leq k' \leq k} d.move[j'+1]  \wedge \neg d.loc.press[j']) 
			\wedge  \bigvee_{i\leq i'\leq k} d.loc.press[i']
	\end{align*}
 
\subsection{Candidate Skolem functions of Case Study}
\label{appendix:skolem}
 \begin{align*}
 \small
	 f.pick[m] \leftrightarrow & \big((\neg \bigvee\limits_{i=0}^{m-1}blank[i] \vee f.loc.feed[m-1] ) \wedge \beta\big) \wedge \big((f.loc.feed[m-1] \wedge \neg i.picked[m-1])\vee \beta \big)\\
	 f.move[m]  \leftrightarrow &  \big((
	 \bigvee\limits_{i=0}^{m}(f.loc.feed[i] \wedge i.picked[i]) \wedge \bigvee\limits_{i=0}^{m}f.loc.press[i]
	\wedge \neg f.loc.feed[m] 
	 \wedge \neg f.loc.press[m]\\
	 & \wedge f.loc.press[m-1]) 
	 \vee ( \bigvee\limits_{i=0}^{m}(f.loc.feed[i] \wedge i.picked[i]) \wedge \bigvee\limits_{i=0}^{m}f.loc.press[i] \wedge \neg f.loc.feed[m] \\
	 & \wedge \neg f.loc.press[m] \wedge i.picked[m]) \vee \beta)\big)
\\
	press[m]  \leftrightarrow &  
	  (i.loc.press[m-1] 
	  \vee  \beta) 
\\
	  d.pick[m] \leftrightarrow &\big( (\neg \bigvee\limits_{i=0}^{m}  i.loc.pressed[i-1]  \vee d.loc.press[m])  \wedge \beta\big)\\	  
	 d.move[m] \leftrightarrow &\big( ((\bigvee\limits_{i=0}^{m-1}(i.picked[i]\wedge i.pressed[i])) \wedge \neg d.loc.deposit[m])\vee \beta\big) \wedge \big((\neg d.loc.deposit[m-1] \\
	 &\vee \neg i.picked[m-1])\wedge \beta \big) \wedge \big( (\bigvee\limits_{i=0} (\neg i.picked[i]\wedge d.loc.deposit[i]))\wedge \neg d.loc.press[m-1]\big)
	 \end{align*}

\subsection{Formulas after Skolemsize for Case Study }
\label{appendix:skolemsize}
$$
 \scriptsize
  \begin{aligned}
		 	&   \bigg( (\neg blank[j] \vee f.loc.feed[j+1]) \bigvee \bigg( \bigwedge_{j\leq j'< k_{1}-1} \Big( \neg f.loc.feed[j'+1] \wedge \neg \big((\neg \bigvee\limits_{i=0}^{m} blank[i]  \vee f.loc.feed[j'+1]) \wedge \beta\big) \\
		 	& \wedge f.loc.feed[k_1] \bigg) \bigg)
 \bigwedge ( \neg f.loc.feed[j] \vee  i.picked[j]  \vee  \big((\neg \bigvee\limits_{i=0}^{j-1}blank[i]  \vee f.loc.feed[j+1]) \wedge \neg i.picked[j]\\
 &
  \wedge f.loc.feed[m] \wedge \beta)\big)
\bigwedge \bigg(\big(\neg f.loc.feed[j] \vee \neg i.picked[j]\vee ((f.loc.press[j+1]\wedge press\_ready[j+1])) \big) \bigvee 
\\
& \bigg( \bigwedge_{j\leq j'< k_{2}-1}\Big( (\neg f.loc.press[j'+1]\vee 
\neg press\_ready[j'+1]) \wedge \big(\neg(f.loc.press[j'] \wedge i.picked[j'] ) \wedge \beta\big)\Big)\bigg)\wedge ( f.loc.press[k_2] \\
&
\wedge   press\_ready[k_2]) \bigg)
\bigwedge
( \neg f.loc.press[j] \vee f.loc.feed[j+1]) \bigvee\bigg(
\bigwedge_{j\leq j'< k_{3}-1}\Big( 
\neg f.loc.feed[j'] \wedge \big(\neg(f.loc.press[j'+1] 
\\
&\wedge i.picked[j'] ) \wedge \beta\big)\Big)
\Big)
\wedge f.loc.feed[k_3]
 \bigg)
\bigwedge\bigg( (\neg f.loc.press[j] \vee \neg i.picked[j]) \bigvee \neg  \big(\neg(f.loc.press[j] \wedge i.picked[j] ) \wedge \beta\big)
 \bigg)
	  \bigwedge
	  \\
	   &\bigg( 
	   (\neg f.loc.press[j] \vee \neg i.picked[j]) \bigvee
((f.loc.press[j] \wedge  i.picked[j] ) 
	  \vee \beta)
	   \bigg)\bigwedge
 \bigg( \neg i.loc.pressed[j] \bigvee (i.loc.press[j] 
	  \vee  \beta)   \bigg)
 \bigwedge \\
 &\bigg( 
 (\neg i.loc.pressed[j] \vee d.loc.press[j+1]) \bigvee 
 \big( 
 \bigwedge_{j\leq j'< k_{4}-1} 
 (
 \neg d.loc.press[j'+1] \wedge \neg\big( (\neg  i.loc.pressed[j']  \vee d.loc.press[j'+1])  \wedge \beta\big) \\
& \wedge  d.loc.press[k_4]
 \bigg)
 \bigwedge  \bigg( 
 \neg d.loc.press[j] \vee i.picked[j] \vee \big( (\neg  i.loc.pressed[j]  \vee d.loc.press[j+1])  \wedge \beta\big)
 \bigg)
 \bigwedge  \bigg( 
 (\neg i.picked[j] \\
 &\vee \neg i.loc.pressed[j] \vee d.loc.deposit[j+1]) \bigvee \big( \bigwedge_{j\leq j'< k_{5}-1} 
 (\neg d.loc.deposit[j'+1] 
 \wedge \big((\neg i.picked[j']  \vee \neg d.loc.deposit[j']) \wedge \beta)  \\
 &
	 \big) \wedge d.loc.deposit[k_5]
 \bigg)\bigwedge\bigg(\neg d.loc.deposit[j] \vee \neg i.picked[j] \vee \neg \big((\neg i.picked[j]  \vee \neg d.loc.deposit[j]) \wedge \beta)
	 \big)\bigg) 
	 \bigwedge\\
	 &\bigg( 
	 \neg d.loc.deposit[j] \vee  \neg i.picked[j] \vee ((d.loc.deposit[j] \wedge i.picked[j]) \vee \beta)
	 \bigg)
	 \bigwedge\\
	 &\bigg( 
	 (i.picked[j]\vee \neg  d.loc.deposit[j] \vee d.loc.press[j+1]) \bigvee \big( 
 \bigwedge_{j\leq j'< k_{6}-1}(\neg  d.loc.deposit[j'+1] \wedge \big((\neg i.picked[j']  \\
& \vee \neg d.loc.deposit[j']) \wedge \beta)
	 \big)) \wedge  d.loc.press[k_6]\big)
	 \bigg)
\end{aligned}
$$ 

\subsection{Unrealizable Core of Case Study}
\label{appendix:unsat}
$$ 
 \scriptsize\begin{aligned}
	& \bigg( (\bigwedge_{j\leq j'< k_{1}-1}(\neg blank[j] \vee f.loc.feed[j+1] \vee \neg f.loc.feed[j'+1] \vee i.picked[j'] \vee  \neg f.loc.feed[j'+1] )) \wedge  
	 \\
	& \wedge (\neg blank[j] \vee f.loc.feed[j+1] \vee  f.loc.feed[k_1]) \bigg) \bigwedge  \bigg( 
	(\neg \bigvee\limits_{i=0}^{j-1}blank[i]  \vee f.loc.feed[j+1]) \wedge \neg i.picked[j]\wedge f.loc.feed[j] )\bigg) \\
	&\bigwedge (\neg i.loc.pressed[j] \vee d.loc.press[j+1]) \vee \neg d.loc.press[j+2] )  \wedge\\
	& (\neg i.loc.pressed[j] \vee d.loc.press[j+1] \vee i.loc.pressed[j+1]) \wedge \cdots \wedge (\neg i.loc.pressed[j] \vee d.loc.press[j+1]) \vee \neg d.loc.press[k_4-1] ) \\
	&	 \wedge(\neg i.loc.pressed[j] \vee d.loc.press[j+1] \vee i.loc.pressed[k_4-2]) \wedge (\neg i.loc.pressed[j] \vee d.loc.press[j+1] \vee d.loc.press[k_4]) \\
	&\bigwedge \bigg( 
 \neg d.loc.press[j] \vee i.picked[j] \vee \neg  i.loc.pressed[j]  \vee d.loc.press[j+1]
 \bigg)\\
 & \bigwedge_{j\leq j'< k_{5}-1}(\neg (i.picked[j]\wedge i.pressed[j]) \vee \neg d.loc.deposit[j'+1] ) \wedge \bigwedge_{j\leq j'< k_{5}-1}(\neg (i.picked[j]\wedge i.pressed[j]) \vee \\
 & \neg (d.loc.deposit[j'-1]\wedge i.picked[j'-1]) )  \wedge (\neg (i.picked[j]\wedge i.pressed[j]) \vee (\bigvee\limits_{i=0}^{j'-1} (\neg i.picked[i]\wedge d.loc.deposit[i])) \wedge \\
 &\bigwedge_{j\leq j'< k_{5}-1}(\neg (i.picked[j]\wedge i.pressed[j])\vee \neg d.loc.press[j'-1]) \wedge \bigwedge_{j\leq j'< k_{5}-1}(\neg (i.picked[j]\wedge i.pressed[j])\vee \neg d.loc.press[j'])\\
 &\wedge (\neg (i.picked[j]\wedge i.pressed[j]) \vee d.loc.deposit[k_5])\\
\end{aligned}$$

\begin{figure}[t]
    \centering
    \includegraphics[width=\linewidth]{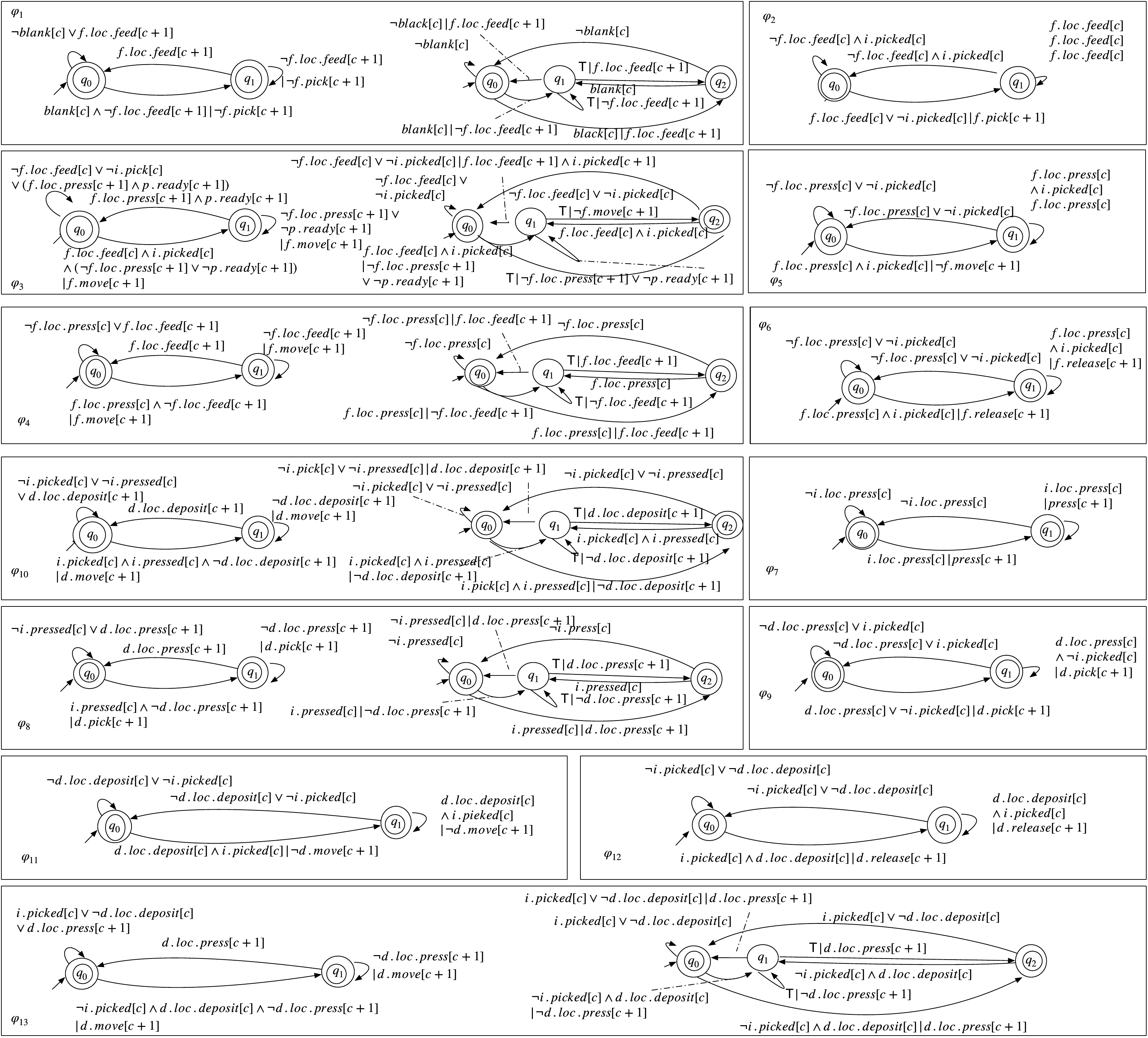}
    \caption{Synthesized production cell controller.}
    \label{fig:robot.gxu}
\end{figure}

\end{document}